\documentclass[letterpaper,twocolumn,prl,aps,superscriptaddress,amsmath,amssymb,floatfix]{revtex4-2}
\usepackage{mathptmx}
\usepackage[latin9]{inputenc}
\usepackage{color}
\usepackage{amsmath}
\usepackage{amssymb}
\usepackage{graphicx}
\usepackage{esint}
\usepackage[unicode=true,
 bookmarks=true,bookmarksnumbered=false,bookmarksopen=false,
 breaklinks=false,pdfborder={0 0 1},backref=false,colorlinks=true]
 {hyperref}
\hypersetup{
 linkcolor=magenta,urlcolor=blue,citecolor=blue,pdfstartview={FitH},hyperfootnotes=false}

\makeatletter

\usepackage{textcomp}
\usepackage{epstopdf}

\pdfpageheight\paperheight
\pdfpagewidth\paperwidth

\@ifundefined{textcolor}{}{%
 \definecolor{BLACK}{gray}{0}
 \definecolor{WHITE}{gray}{1}
 \definecolor{RED}{rgb}{1,0,0}
 \definecolor{GREEN}{rgb}{0,1,0}
 \definecolor{BLUE}{rgb}{0,0,1}
 \definecolor{CYAN}{cmyk}{1,0,0,0}
 \definecolor{MAGENTA}{cmyk}{0,1,0,0}
 \definecolor{YELLOW}{cmyk}{0,0,1,0}
}

\usepackage{xcolor}\usepackage{soul}
\newcommand{\ket}[1]{\ensuremath{\left|#1\right\rangle}}

\definecolor{blue}{rgb}{0,0,1}
\definecolor{red}{rgb}{1,0,0}
\definecolor{green}{rgb}{0,1,0}

\usepackage{soul}

\makeatother

\begin{document}
\title{Cooperative quantum interface for noise mitigation in quantum networks}
\author{Yan-Lei Zhang}
\thanks{These two authors contributed equally to this work.}
\affiliation{Key Laboratory of Quantum Information, CAS, University of Science
and Technology of China, Hefei 230026, China}
\affiliation{CAS Center For Excellence in Quantum Information and Quantum Physics,
University of Science and Technology of China, Hefei, Anhui 230026,
P. R. China}
\author{Ming Li}
\thanks{These two authors contributed equally to this work.}
\affiliation{Key Laboratory of Quantum Information, CAS, University of Science
and Technology of China, Hefei 230026, China}
\affiliation{CAS Center For Excellence in Quantum Information and Quantum Physics,
University of Science and Technology of China, Hefei, Anhui 230026,
P. R. China}
\author{Xin-Biao Xu}
\affiliation{Key Laboratory of Quantum Information, CAS, University of Science
and Technology of China, Hefei 230026, China}
\affiliation{CAS Center For Excellence in Quantum Information and Quantum Physics,
University of Science and Technology of China, Hefei, Anhui 230026,
P. R. China}
\author{Chun-Hua Dong}
\affiliation{Key Laboratory of Quantum Information, CAS, University of Science
and Technology of China, Hefei 230026, China}
\affiliation{CAS Center For Excellence in Quantum Information and Quantum Physics,
University of Science and Technology of China, Hefei, Anhui 230026,
P. R. China}
\affiliation{Hefei National Laboratory, University of Science and Technology of
China, Hefei 230088, China.}
\author{Guang-Can Guo}
\affiliation{Key Laboratory of Quantum Information, CAS, University of Science
and Technology of China, Hefei 230026, China}
\affiliation{CAS Center For Excellence in Quantum Information and Quantum Physics,
University of Science and Technology of China, Hefei, Anhui 230026,
P. R. China}
\affiliation{Hefei National Laboratory, University of Science and Technology of
China, Hefei 230088, China.}
\author{Ze-Liang Xiang}
\email{xiangzliang@mail.sysu.edu.cn}

\affiliation{School of Physics, Sun Yat-sen University, Guangzhou 510275, China.}
\author{Chang-Ling Zou}
\email{clzou321@ustc.edu.cn}

\affiliation{Key Laboratory of Quantum Information, CAS, University of Science
and Technology of China, Hefei 230026, China}
\affiliation{CAS Center For Excellence in Quantum Information and Quantum Physics,
University of Science and Technology of China, Hefei, Anhui 230026,
P. R. China}
\affiliation{Hefei National Laboratory, University of Science and Technology of
China, Hefei 230088, China.}
\author{Xu-Bo Zou}
\email{xbz@ustc.edu.cn}

\affiliation{Key Laboratory of Quantum Information, CAS, University of Science
and Technology of China, Hefei 230026, China}
\affiliation{CAS Center For Excellence in Quantum Information and Quantum Physics,
University of Science and Technology of China, Hefei, Anhui 230026,
P. R. China}
\affiliation{Hefei National Laboratory, University of Science and Technology of
China, Hefei 230088, China.}
\date{\today}
\begin{abstract}
Quantum frequency converters that enable the interface between the itinerant photons and qubits are indispensable for realizing long-distance quantum network. However, the cascaded connection between converters and qubits usually brings additional insertion loss and intermediate noises. Here, we propose a cooperative quantum interface (CQI) that integrates the converter and qubit coupling into a single device for efficient long-distance entanglement generation. Compared to traditional cascaded systems, our scheme offers several advantages, including compactness, reduced insertion loss, and suppression of noise from intermediate modes.  We prove the excellent performance over the separated devices by about two orders of magnitude for the entangled infidelity of two remote nodes. Moreover, we discuss an extended scheme for multiple remote nodes, revealing an exponential advantage in performance as the number of nodes increases.  The cooperative effect is universal that can be further applied to multifunctional integrated quantum devices. This work opens up novel prospects for quantum networks, distributed quantum computing, and sensing.
\end{abstract}
\maketitle
\emph{Introduction.-} Quantum networks \citep{kimble2008quantum,wehner2018quantum,Reiserer2022,Azuma2023}
promise to revolutionize information processing by enabling secure communication~\cite{Langenfeld2021}, distributed quantum computing~\cite{Jiang2007}, and enhanced sensing capabilities~\cite{Malia2022,Guo2020}. The realization of quantum networks relies on the efficient integration of quantum information processors with quantum communication channels. Quantum interfaces play a crucial role in coherently exchanging information and generating entanglement between local memory matter qubits across diverse physical platforms, including atoms~\citep{chou2005measurement,hofmann2012heralded},
trapped ions~\citep{moehring2007entanglement}, superconducting qubits~\citep{kurpiers2018deterministic}, quantum dots~\citep{delteil2016generation},
and spin impurities~\citep{humphreys2018deterministic}.
In particular, long-distance quantum communications~\citep{yuan2010entangled,hensen2015loophole, Zhou2024long},
employ itinerant photons~\citep{kurpiers2018deterministic} as information carriers, leveraging their low-loss transmission through waveguide~\citep{gonzalez2011entanglement},
optical fiber~\citep{inagaki2013entanglement,yu2020entanglement}, or free space~\citep{yin2017satellite,huang2024}. However, a significant challenge arises from the frequency mismatch between memory qubits, often operating in the visible or microwave range, while the telecommunication wavelengths are optimal for long-distance transmission~\citep{yu2020entanglement}.

Quantum frequency conversion (QFC)~\citep{kumar1990quantum} has emerged as a promising solution to bridge this gap, enabling the coherent quantum state transfer between disparate frequency domains. For example, nonlinear optical crystals have been used to convert photons from the visible to telecommunication  wavelengths~\cite{ikuta2011wide,Krutyanskiy2024,Zhou2024}. Quantum transducers, such as the optomechanical systems and electro-optic modulators, have been employed to mediate the coupling between the spin or superconducting qubits and optical photons through intermediate microwave photons or phonons \citep{rabl2010quantum, benito2020hybrid, mirhosseini2020superconducting, wallquist2009hybrid, kurizki2015quantum,Han2021,Balram2022}. Recent experimental successes have brought the realization of long-distance quantum networks closer to reality~\citep{liu2024creation,Knaut2024,Stolk2024}.  Despite these advances, significant challenges remain. Current approaches often require multiple cascaded conversion steps for each network node, typically involving bidirectional QFC between telecom photons and intermediate carriers for coupling the memory qubits~\citep{yu2020entanglement, van2022entangling}. This cascaded architecture introduces insertion losses and additional noise from intermediate modes, limiting the overall efficiency and fidelity of quantum interfaces.


\begin{figure}
\includegraphics[width=0.5\textwidth]{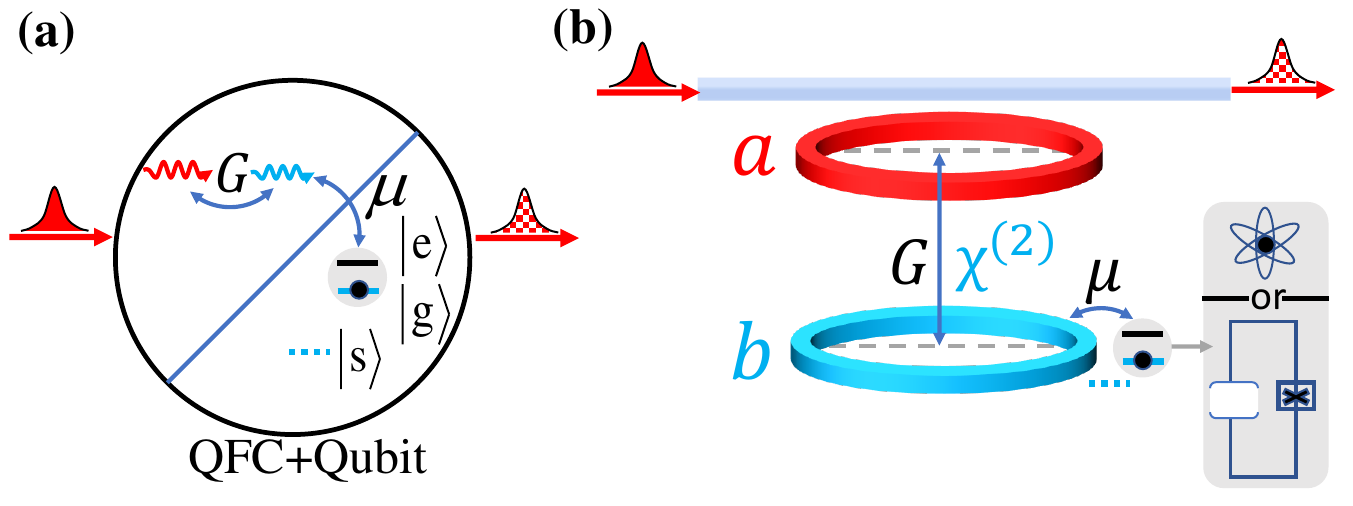}
\caption{Concept and implementation of the cooperative quantum interface (CQI). (a) The hybrid CQI device integrates quantum frequency conversion (QFC) and qubit coupling into a single unit. The CQI enables direct entanglement generation between an input telecom photon and a memory qubit, while suppressing noise from intermediate modes. (b) Potential experimental realization of the CQI. The coupling between telecom photon modes $(a)$ and intermediate mode $(b)$ can be achieved through optical $\chi^{\left(2\right)}$ nonlinearity or optomechanical couplings. The qubit can be implemented using atoms, superconducting circuits, or quantum dots, and the intermediate mode corresponds to an visible optical, microwave or phonon mode, depending on the specific experimental system.}
\label{Fig1}
\end{figure}

To address these challenges, we propose a cooperative quantum interface (CQI) that integrates QFC and qubit coupling into a single device. The CQI leverages the cooperative effect to significantly suppress noise from intermediate modes. Our scheme demonstrates a significant improvement in the fidelity and efficiency of remote entanglement generation, with the infidelity reduced by nearly two orders of magnitude compared to traditional cascaded approach.
Remarkably, this advantage becomes even more pronounced when scaling to multiple nodes, with the comparative overall performance increasing exponentially with the number of nodes. By harnessing the hybrid vigor of CQIs, our work opens up new avenues for the realization of scalable and high-performance quantum networks, and also inspires novel designs of high-performance functional devices with hybrid quantum systems.

\emph{Model.-} Figure~\ref{Fig1}(a) illustrates the scheme of a hybrid CQI that seamlessly integrates multiple physical processes into a single device . This integration offers two significant advantages over conventional cascaded approaches. On one hand, it eliminates insertion losses typically associated with intermediate excitation modes. On the other hand, it enhances the efficiency by a factor of $1/\eta_{\mathrm{FC}}^{2}$, where $\eta_{\mathrm{FC}}$ is the efficiency of an isolated QFC device (see Supplementary Material~\cite{SM} for detailed derivations). Figure~\ref{Fig1}(b) presents potential experimental realizations of the CQI, where the optical mode $(a)$ and intermediate mode $(b)$ are depicted as two coupled rings, which could be two modes in a single microring resonator~\cite{Wang2021, ilchenko2004nonlinear} or two coherently coupled distinct resonators~\cite{Javerzac-Galy2016,fan2018superconducting}. A memory matter qubit is coupling to the mode $b$, completing the hybrid interface. The interaction Hamiltonian of this integrated system is given by
\begin{equation}
H_{\mathrm{int}} = G\left(a^{\dagger}b+ab^{\dagger}\right)+\mu\left(\sigma_{+}b+\sigma_{-}b^{\dagger}\right),
\label{Eq1}
\end{equation}
where $G$ and $\mu$ are the coupling strengths, $a$ and $b$ are the annihilation
operators for the telecom optical mode and intermediate mode, respectively, and $\sigma_{+}=\left|e\right\rangle \left\langle g\right|$ is the transition operator of the qubit. This Hamiltonian encapsulates two processes: the first term describes the QFC process, and the second term enables the quantum gate between the qubit and the intermediate carrier. Crucially, our CQI incorporate both terms into a single device, in contrast to conventional cascaded approaches where these processes are implemented in separate devices, requiring a sequence of QFC$\rightarrow$qubit interaction$\rightarrow$QFC to  interface between input telecom photon ($a$) and the memory qubit states $\mathrm{span}\{\ket{g},\ket{s}\}$.

When a single telecom photon is sent to CQI, the quantum state $\left|\Psi\left(t\right)\right\rangle$ describing the entire system can be expanded in the single-excitation subspace~\citep{shen2005coherent, rosenblum2017analysis, yan2023unidirectional}. The dynamics of the system are governed by the Schrodinger equation $i\frac{\partial}{\partial t}\left|\Psi\left(t\right)\right\rangle =H_{\mathrm{sys}}\left|\Psi\left(t\right)\right\rangle$, where $H_{\mathrm{sys}}$ is the Hamiltonian of the entire system. Through the CQI, the entanglement is established between the output photon (reflected or transmitted) and the qubit. The entanglement serves as a building block for generating entanglement between remote nodes via state transfer or entanglement swapping protocols~\citep{pan1998experimental}. Analytical solutions for the single-excitation subspace can be obtained~\cite{SM}, with the complex amplitude of the output photon being
\begin{equation}
f_{\mu}=1-\frac{2\kappa_{a,ex}/\kappa_{a}}{1+C_{ab}/\left(1+C_{bq}\right)}
\end{equation}
for the on-resonance condition. Here, $C_{bq}=4\mu^{2}/\left(\gamma\kappa_{b}\right)$ and $C_{ab}=4G^{2}/\left(\kappa_{a}\kappa_{b}\right)$ are the cooperativities, $\kappa_{a(b)}=\kappa_{a(b),o}+\kappa_{a(b),ex}$ is the total decay rate of mode $a(b)$ with the intrinsic loss $\kappa_{a(b),o}$ and the external coupling rate $\kappa_{a(b),ex}$, and $\gamma$ is the qubit decay rate. In the ideal case, with negligible intrinsic photon loss ($\kappa_{a,ex}=\kappa_{a}$) and strong qubit-intermediate mode coupling $C_{bq}\gg C_{ab}$, we obtain $f_{\mu}\approx-1$. When the qubit is in the auxiliary state $\left|s\right\rangle $, there is no interaction with the qubit ($\mu=0$), yielding $f_{\mu=0}\approx1$ for $C_{ab}\gg1$. Thus, the output photon acquires the opposite phase when the qubit is in the state $\left|g\right\rangle $ or $\left|s\right\rangle $, which enables the realization of a controlled-phase gate and generate entanglement between the photon and qubit~\cite{Duan2004}. This fundamental operation forms the basis for more complex quantum network protocols and distributed quantum information processing tasks.

\begin{figure}
\includegraphics[width=0.5\textwidth]{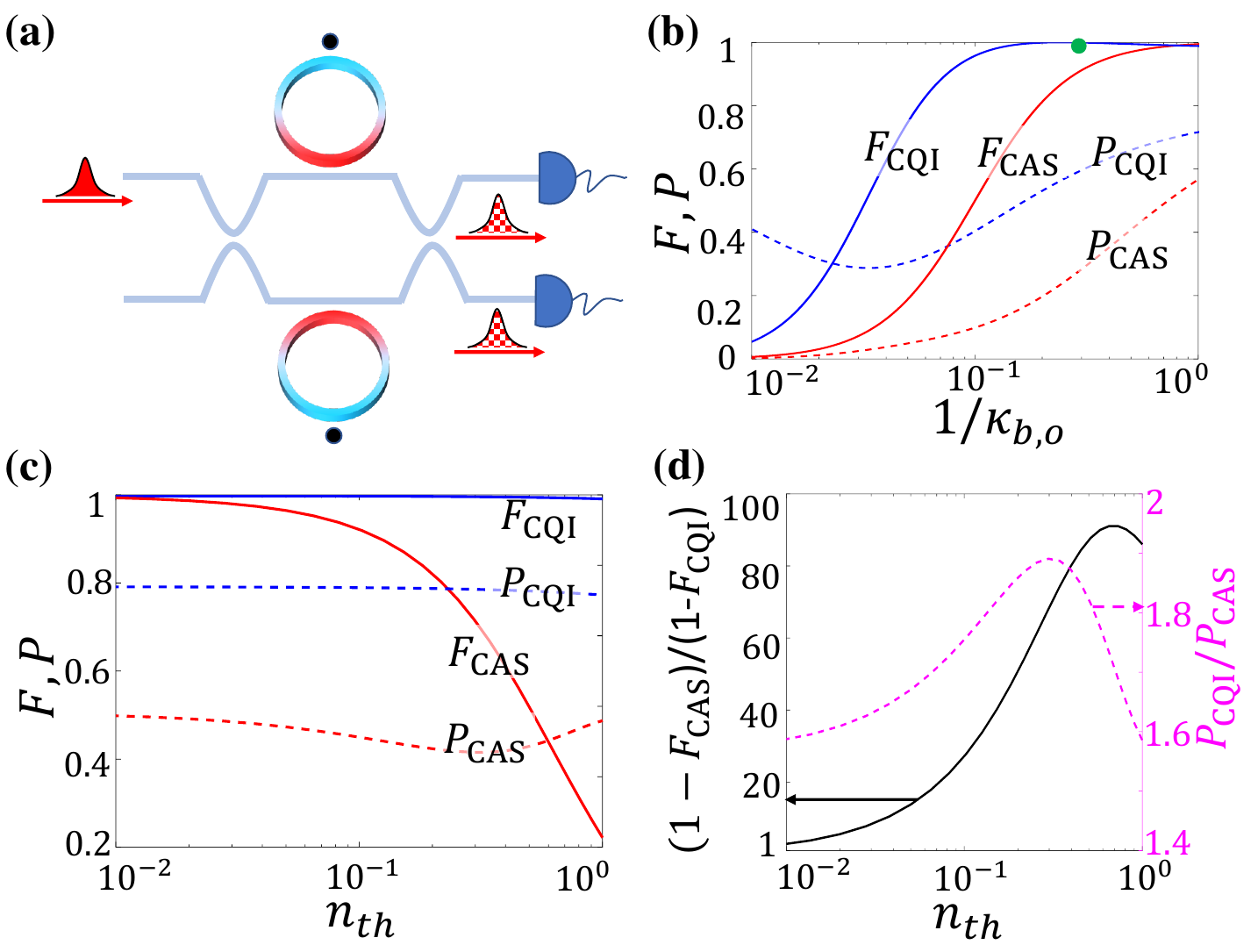}
\caption{(a) Scheme of the entanglement preparation between two remote nodes using a CQI at each node. (b) Comparison between the CQI and the traditional cascaded (CAS) scheme in the fidelity $F$ and the successful $P$ versus the intrinsic lifetime $1/\kappa_{b,o}$ of the intermediate mode $b$. Green dot: the ideal case where $F_{\mathrm{CQI}}=1$. (c) and (d) Noise resilience of CQIs from intermediate mode $b$ with a thermal equilibrium excitation number $n_{th}$. The quantitative superiority of CQI over CAS are also shown by the infidelity ratio $\left(1-F_{\mathrm{CAS}}\right)/\left(1-F_{\mathrm{CQI}}\right)$ and the efficiency ratio $P_{\mathrm{CQI}}/P_{\mathrm{CAS}}$. Parameters: $\kappa_{a,o}=1$, $\kappa_{a,ex}=14$, and $G=\mu=10$ (all normalized to the qubit decay rate $\gamma$), $\kappa_{b,ex}=10$ for CQI. In (c) and (d), the dissipation rate is $\kappa_{b,o}=0.1$}
\label{Fig2}
\end{figure}

\emph{Entanglement of two remote nodes.-} A simple scheme for generating entanglement between two remote nodes is plotted in Fig.~\ref{Fig2}(a). A single telecom photon passes through a beam splitter, resulting in a superposition state of the photon in the upper ($\ket{\uparrow}$) and lower ($\ket{\downarrow}$) path, i. e. $\ket{\pm i}=\left(\left|\uparrow\right\rangle \pm i\left|\downarrow\right\rangle \right)/\sqrt{2}$ . After travelling a long distance through the fiber, the photon in each path couples with the qubit $1$ and qubit $2$ through the CQI, which are represented by colored microrings. Let the initial state of the overall system be
\begin{equation}
\left|\Psi\right\rangle _{0} =\frac{1}{2}\ket{+i}\otimes\left(\left|s\right\rangle _{1}+\left|g\right\rangle _{1}\right)\otimes\left(\left|s\right\rangle _{2}+\left|g\right\rangle _{2}\right).
\end{equation}
Ignoring photon loss and other imperfections, the ideal final state is given by
\begin{eqnarray}
\left|\Psi\right\rangle_{\mathrm{ideal}} & = & \frac{1}{2}\ket{+i}\left(\left|s\right\rangle _{1}\left|s\right\rangle _{2}-\left|g\right\rangle _{1}\left|g\right\rangle _{2}\right)\nonumber \\
 &  & +\frac{1}{2}\ket{-i}\left(\left|s\right\rangle _{1}\left|g\right\rangle _{2}-\left|g\right\rangle _{1}\left|s\right\rangle _{2}\right).
 \label{Eq4}
\end{eqnarray}
As shown in Fig.~\ref{Fig2}(a), by measuring the output photon state in the basis of $\ket{\pm i}$ through the beam splitter and single photon detectors,  the two qubits will be projected into entangled Bell states.  The post-selection measurement,  where only a specific measurement result is selected and the other results are discarded, is a common and  highly effective  method to implement  the entanglement, quantum  cryptography, and quantum metrology \citep{hofmann2002quantum, christandl2009postselection, arvidsson2020quantum}.
By applying further adaptive single qubit operation $Z$ or $Y$ gates to qubit 1 or qubit 2 according to the measurement outcome, the Bell state  $\left(\left|s\right\rangle _{1}\left|s\right\rangle _{2}+\left|g\right\rangle _{1}\left|g\right\rangle _{2}\right)/\sqrt{2}$ can be generated deterministically.

In practical experimental systems, various noises must be considered.  To quantify the performance of the devices, we define the fidelity of the final state $\left|\Psi\right\rangle_{\mathrm{f}}$ under realistic noise conditions compared with the ideal state as $F=\left|\left\langle \Psi\right|_{\mathrm{f}}\left|\Psi\right\rangle _{\mathrm{ideal}}\right|^{2}/P$ with $P=\left\langle \Psi\right|_{\mathrm{f}}\left|\Psi\right\rangle_{\mathrm{f}}$ being the normalization factor that also denotes the success probability of detecting the output photon.  Analytical analysis reveals that the fidelity and successful probability are fully determined by the complex amplitude of the output photon from the CQI, which can be expressed as~\cite{SM} $F_{\mathrm{CQI}}=\left|f_{\mu=0}-f_{\mu}\right|^{2}/4P_{\mathrm{CQI}}$ and $P_{\mathrm{CQI}}=\left(\left|f_{\mu=0}\right|^{2}+\left|f_{\mu}\right|^{2}\right)/2$. Figure \ref{Fig2}(b) depicts the performance versus the reciprocal of decay rate $1/\kappa_{b,o}$ of the intermediate mode under resonant condition. $F_{\mathrm{CQI}}$ increases as $\kappa_{b,o}$ decreases, and the green dot indicate the ideal output with $F_{\mathrm{CQI}}=1$ when the analogous impedance matching in the CQI is achieved~\cite{SM}. For comparison, the fidelity and successful probability (labelled by $F_\mathrm{CAS}$ and $P_{\mathrm{CAS}}$) of the cascaded  approach  are plotted with red lines. The model of cascaded approach has all parameters as the same as the CQI, while three $b$ modes are introduced for three individual devices for the forth and back QFCs and the cavity-qubit coupling, with an external coupling rate $\kappa_{b,ex}$ between each $b$ mode to a bus channel connecting the devices. In addition, we also notice that the QFC can be realized by the perlodlcally poled lithlum nlobate waveguide~\cite{van2022entangling}, and we can not make a direct comparison due to different matching conditions and optimal parameters. However, at least in terms of interation, the cascaded devices are far less than the CQI.
These results clearly demonstrate the advantages of the CQI in terms of fidelity, efficiency, or integration.

\begin{figure}
\includegraphics[width=0.5\textwidth]{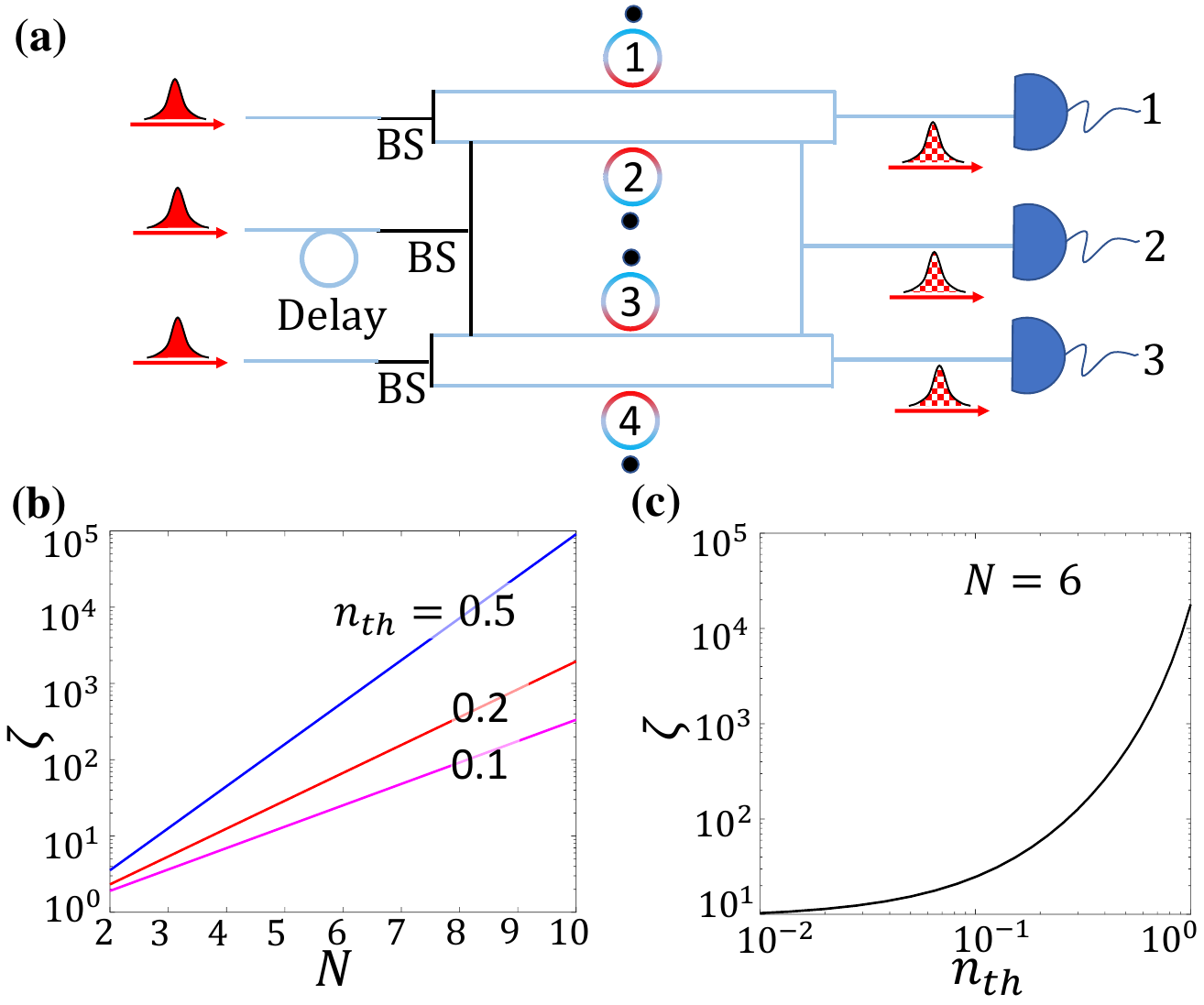}
\caption{(a) An expended scheme of four remote nodes. BS:  beam splitter. The CQI shows the huge advantage by the enhancement ratio
$\zeta$ with the increasing nodes $N$ (b), where $\zeta=F_{\mathrm{CQI}}P_{\mathrm{CQI}}/\left(F_{\mathrm{CAS}}P_{\mathrm{CAS}}\right)$
is the effective factor that contains the fidelity and the successful probability.
It is obvious that the performance is better for more thermal noise,
and the black line (c) for the $N=6$ nodes further confirms the result.
The selection of relevant parameters is the same as the Fig. \ref{Fig2}.
}
\label{Fig3}
\end{figure}

In addition to vacuum noise due to intrinsic loss, thermal noise from the intermediate mode $b$ is inevitable in the low-frequency band, such as the the microwave photon or phonon mode in hybrid superconducting system~\citep{xiang2013hybrid,clerk2020hybrid}. Considering the thermal noises, the evolution of the full system can be solved by a master equation~\cite{SM}, and our numerical results are confirmed with the analytical results in the absence of noise. Figure \ref{Fig2}(c) shows the noise immunity for the intermediate process from weak to strong thermal noises, with $n_{th}$ denoting the mean thermal excitation number of mode $b$. In the  cascaded approach, the output photon from mode $a$ is greatly influenced by the noise from mode $b$, regardless of whether the noise originates from the QFC process or the functional device. These findings are confirmed by the numerical results (red lines) in Fig. \ref{Fig2}(c). For the CQI, the maximal fidelity does not necessarily occur at the resonance point, particularly in noisy situations, due to the cooperative effect.

The underlying mechanism can be explained with the supermodes representation  $(a\mp b)/\sqrt{2}$  with the corresponding supermode frequency according to the QFC interaction term $G\left(a^{\dagger}b+ab^{\dagger}\right)$. Although the photon loss can be also addressed through post selection for both the cascaded scheme and CQI, the acquired phases that are related to the entanglement fidelity are different for the two schemes.
For relatively large noise, the effect from the intermediate process is partially isolated from the combined system around the supermode frequency ~\cite{SM}. Unlike linearly coherent transduction~\citep{wang2022generalized} by employing dark modes~\citep{wang2012using}, our model is nonlinear and any other excitation is not needed other than the driven mode, which leads that the noise from the  intermediate mode can be isolated by controlling the driving detuning.
Therefore we can obtain the optimal fidelity $F_{\mathrm{CQI}}$ (blue lines) by tuning the driving frequency of the input photon. To demonstrate this advantage quantitatively, we calculate the infidelity ratio $\left(1-F_{\mathrm{CAS}}\right)/\left(1-F_{\mathrm{CQI}}\right)$ and the efficiency ratio $P_{\mathrm{CQI}}/P_{\mathrm{CAS}}$, as shown in Fig. \ref{Fig2}(d). The numerical results indicate that the infidelity ratio could be improves by nearly two orders of magnitude when the thermal excitation is not negligible, and the efficiency always surpasses that of the cascaded system.

\emph{Cooperative advantage of multiple nodes.-} Quantum networks require the realization of entanglement among multiple remote nodes. Our scheme demonstrates significant advantages compared to the conventional cascaded system, which requires more QFC devices and introduces more insertion losses and noises, and greatly limits the extension of the quantum network to more nodes. Therefore, the CQI in our scheme offers potential advantages for expanding remote nodes to construct quantum networks. As an example, the scheme to expand two nodes to four nodes is shown in Fig. \ref{Fig3}(a), and the expansion to even more nodes could be realized by iterating the scheme. Initially, we prepare two pairs of entangled states with two nodes each, which can be written as $\left|\Psi\right\rangle _{1,2}=\left(\left|s\right\rangle _{1}\left|s\right\rangle _{2}+\left|g\right\rangle _{1}\left|g\right\rangle _{2}\right)/\sqrt{2}$ and $\left|\Psi\right\rangle _{3,4}=\left(\left|s\right\rangle _{3}\left|s\right\rangle _{4}+\left|g\right\rangle _{3}\left|g\right\rangle _{4}\right)/\sqrt{2}$. This setp requires two single photon inputs and measurements. Subsequently, a third single photon is prepared to the superposition of two paths $\ket{+i}$ for  interacting with the qubit $2$ and $3$, respectively. A time delay is introduced to ensure that there are no more than two photon excitation presents in each path simultanesly. Following the same procedure as establishing entanglement between two nodes [Eq.~{\ref{Eq4})], the detection of the output photon in $\ket{\pm i}$ projects the 4 qubits into $\left( \ket{s}_1\ket{s}_2\ket{s}_3\ket{s}_4 -\ket{g}_1\ket{g}_2\ket{g}_3\ket{g}_4 \right)/2$ and $\left( \ket{s}_1\ket{s}_2\ket{g}_3\ket{g}_4 -\ket{g}_1\ket{g}_2\ket{s}_3\ket{s}_4 \right)/2$, and we achieve the four-node Greenberger-Horne-Zeilinger-type entanglement state through further adaptive control according to the measurement outcome~\cite{SM}.

Therefore, three single photons and measurements are needed for four nodes. To scale up to $N$ multiple nodes, $N-1$ single photons and measurements are needed. For simplicity, we assume that the entangled state of any two nodes is same and $f_{\mu=0}\approx-f_{\mu}$. It can be directly concluded that the overall performance is exponentially related to the number of nodes $N$, where we use the fidelity without the post-selection as the performance indicator, i.e., $F\times P$. In Fig. \ref{Fig3}(b), we plot the enhancement ratio of the performance indicator $\zeta=F_{\mathrm{CQI}}P_{\mathrm{CQI}}/F_{\mathrm{CAS}}P_{\mathrm{CAS}}$ as a function of the number of nodes $N$ for different thermal noise levels $n_{th}=0.5$, $0.2$, and $0.1$, and compare it with the cascaded system. The results show that the enhancement of the indicator increases exponentially with $N$, and the enhancement is more pronounced when the thermal noises is stronger. As further confirmed in Fig. \ref{Fig3}(c), the enhancement shows a super-linear increase with $n_{\mathrm{th}}$. This implies that we can build quantum networks with more nodes at the same noise level by CQI.


\emph{Experimental feasibility.-} The versatility and broad applicability of CQIs make them a promising approach for experimental realization. The coupling between the telecom mode $a$  and the intermediate mode $b$, as depicted in Fig.~\ref{Fig1}(b), can be realized through various physical processes, such as the three-wave mixing process based on $\chi^{\left(2\right)}$~\cite{gao2021broadband,lu2020toward}, the four-wave mixing process based on $\chi^{(3)}$  optical nonlinearities~\cite{Vernon2016}, as well as the optomechanical and Brillouin scattering interaction between light and mechanical vibrations~\cite{Han2021,Balram2022,shen2023nonreciprocal}. The choice of the specific implementation will depend on factors such as the achievable cooperativities, the coherence times of the qubits and intermediate modes, and the compatibility with existing quantum communication infrastructure.

One promising experimental scheme involves a neutral atom as the qubit, with the target atomic transition at visible wavelengths, e.g. $780\,\mathrm{nm}$ for Rubidium atoms. In this case, the CQI can be realized using with  a lithium niobate (LN) microring resonator on a photonic chip, which simultaneously support high-quality factor visible ($b$) and telecom ($a$) modes. By leveraging the strong $\chi^{\left(2\right)}$ optical nonlinearity of LN, the QFC from the telecom wavelength to the atomic transition wavelength can be realized through difference-frequency generation process~\cite{lu2021ultralow, marty2021photonic}. Strong coupling between the atom and the visible photon in the resonator can be achieved through the evanescent field, with the atom been trapped in a dipole optical trap and placed to the evanescent field of the cavity mode~\cite{tiecke2014nanophotonic, aoki2006observation, shomroni2014all, scheucher2016quantum,Zhou2024}. Although the thermal noise from intermediate modes can be ignored for this kind of experimental scheme, the CQI can also suppress the dephasing~\cite{panuski2020fundamental}  in addition to the decay of intermediate modes.


Another promising scheme employs superconducting quantum qubit on a hybrid photonic-superconducting circuit platform~\citep{xiang2013hybrid, mirhosseini2020superconducting}. In this case, the intermediate mode $b$ could be a GHz-frequency confined phononic mode in an acoustic cavity, which couples to the qubit through the piezo-mechanical effect~\citep{han2020cavity,Han2021,Balram2022}, and to the telecom photon through the optomechanical interaction~\cite{shen2023nonreciprocal}. This scheme could also be realized with other intermediate modes, such as magnon, microwave photon, and phonon \citep{li2018magnon,lachance2019hybrid,yeo2014strain, li2020enhancing,fan2018superconducting}, given the  experimental demonstrations of coherent QFC between optical photons and these microwave excitations. More discussion of experimental feasibility, such as quantitative comparisons and pulse cases, is in the Supplementary Material~\cite{SM}.

\emph{Conclusion.-} In summary, we have demonstrated the superiority of the CQI  in mitigating intermediate noises and enhancing the performance of remote entanglement generation and node expansion in quantum networks. The CQI exhibits an exponential improvement in overall performance compared to the traditional cascaded method, making it widely applicable to various physical systems prone to unavoidable noises \citep{yu2023telecom, lai2024realization, knaut2024entanglement, chang2020remote, bienfait2019phonon}. Beyond the single photon-qubit interface, the CQI could be employed in photonic quantum error correction codes to enhance their performance and mitigate the impact of noise and the development of noise-resilient quantum memories. Moreover, the concept of the CQI can be generalized to incorporate SU(1,1)-type coupling~\cite{yurke19862}, which allows for the direct generation of qubit-photon entanglement while simultaneously mitigating noise. As we continue to explore the potential of the CQI, we anticipate significant advancements in the realization of scalable and high-performance quantum networks and the emergence of novel devices exploring hybrid quantum system.


\acknowledgements
This work was funded by the National Key R$\&$D Program (Grant No. 2021YFA1402004), the National Natural Science Foundation of China (Grants 92265108, U21A20433, U21A6006, 12104441, 12061131011, 12374361, and 92265210). This work was also supported by the Fundamental Research Funds for the Central Universities and USTC Research Funds of the Double First-Class Initiative. The numerical calculations in this paper have been done on the supercomputing system in the Supercomputing Center of the University of Science and Technology of China. This work was partially carried out at the USTC Center for Micro and Nanoscale Research and Fabrication.

\bibliographystyle{Zou}
\nocite{*}
\bibliography{main}

\begin{thebibliography}{76}%
\makeatletter
\providecommand \@ifxundefined [1]{%
 \@ifx{#1\undefined}
}%
\providecommand \@ifnum [1]{%
 \ifnum #1\expandafter \@firstoftwo
 \else \expandafter \@secondoftwo
 \fi
}%
\providecommand \@ifx [1]{%
 \ifx #1\expandafter \@firstoftwo
 \else \expandafter \@secondoftwo
 \fi
}%
\providecommand \natexlab [1]{#1}%
\providecommand \enquote  [1]{``#1''}%
\providecommand \bibnamefont  [1]{#1}%
\providecommand \bibfnamefont [1]{#1}%
\providecommand \citenamefont [1]{#1}%
\providecommand \href@noop [0]{\@secondoftwo}%
\providecommand \href [0]{\begingroup \@sanitize@url \@href}%
\providecommand \@href[1]{\@@startlink{#1}\@@href}%
\providecommand \@@href[1]{\endgroup#1\@@endlink}%
\providecommand \@sanitize@url [0]{\catcode `\\12\catcode `\$12\catcode
  `\&12\catcode `\#12\catcode `\^12\catcode `\_12\catcode `\%12\relax}%
\providecommand \@@startlink[1]{}%
\providecommand \@@endlink[0]{}%
\providecommand \url  [0]{\begingroup\@sanitize@url \@url }%
\providecommand \@url [1]{\endgroup\@href {#1}{\urlprefix }}%
\providecommand \urlprefix  [0]{URL }%
\providecommand \Eprint [0]{\href }%
\providecommand \doibase [0]{http://dx.doi.org/}%
\providecommand \selectlanguage [0]{\@gobble}%
\providecommand \bibinfo  [0]{\@secondoftwo}%
\providecommand \bibfield  [0]{\@secondoftwo}%
\providecommand \translation [1]{[#1]}%
\providecommand \BibitemOpen [0]{}%
\providecommand \bibitemStop [0]{}%
\providecommand \bibitemNoStop [0]{.\EOS\space}%
\providecommand \EOS [0]{\spacefactor3000\relax}%
\providecommand \BibitemShut  [1]{\csname bibitem#1\endcsname}%
\let\auto@bib@innerbib\@empty
\bibitem [{\citenamefont {Kimble}(2008)}]{kimble2008quantum}%
  \BibitemOpen
  \bibfield  {author} {\bibinfo {author} {\bibfnamefont {H.~J.}\ \bibnamefont
  {Kimble}},\ }\bibfield  {title} {\enquote {\bibinfo {title} {The quantum
  internet},}\ }\href {\doibase 10.1038/nature07127} {\bibfield  {journal}
  {\bibinfo  {journal} {Nature}\ }\textbf {\bibinfo {volume} {453}},\ \bibinfo
  {pages} {1023} (\bibinfo {year} {2008})}\BibitemShut {NoStop}%
\bibitem [{\citenamefont {Wehner}\ \emph {et~al.}(2018)\citenamefont {Wehner},
  \citenamefont {Elkouss},\ and\ \citenamefont {Hanson}}]{wehner2018quantum}%
  \BibitemOpen
  \bibfield  {author} {\bibinfo {author} {\bibfnamefont {S.}~\bibnamefont
  {Wehner}}, \bibinfo {author} {\bibfnamefont {D.}~\bibnamefont {Elkouss}}, \
  and\ \bibinfo {author} {\bibfnamefont {R.}~\bibnamefont {Hanson}},\
  }\bibfield  {title} {\enquote {\bibinfo {title} {Quantum internet: A vision
  for the road ahead},}\ }\href
  {https://www.science.org/doi/full/10.1126/science.aam9288} {\bibfield
  {journal} {\bibinfo  {journal} {Science}\ }\textbf {\bibinfo {volume} {362}}
  (\bibinfo {year} {2018})}\BibitemShut {NoStop}%
\bibitem [{\citenamefont {Reiserer}(2022)}]{Reiserer2022}%
  \BibitemOpen
  \bibfield  {author} {\bibinfo {author} {\bibfnamefont {A.}~\bibnamefont
  {Reiserer}},\ }\bibfield  {title} {\enquote {\bibinfo {title} {{Colloquium:
  Cavity-enhanced quantum network nodes}},}\ }\href {\doibase
  10.1103/RevModPhys.94.041003} {\bibfield  {journal} {\bibinfo  {journal}
  {Reviews of Modern Physics}\ }\textbf {\bibinfo {volume} {94}},\ \bibinfo
  {pages} {041003} (\bibinfo {year} {2022})}\BibitemShut {NoStop}%
\bibitem [{\citenamefont {Azuma}\ \emph {et~al.}(2023)\citenamefont {Azuma},
  \citenamefont {Economou}, \citenamefont {Elkouss}, \citenamefont {Hilaire},
  \citenamefont {Jiang}, \citenamefont {Lo},\ and\ \citenamefont
  {Tzitrin}}]{Azuma2023}%
  \BibitemOpen
  \bibfield  {author} {\bibinfo {author} {\bibfnamefont {K.}~\bibnamefont
  {Azuma}}, \bibinfo {author} {\bibfnamefont {S.~E.}\ \bibnamefont {Economou}},
  \bibinfo {author} {\bibfnamefont {D.}~\bibnamefont {Elkouss}}, \bibinfo
  {author} {\bibfnamefont {P.}~\bibnamefont {Hilaire}}, \bibinfo {author}
  {\bibfnamefont {L.}~\bibnamefont {Jiang}}, \bibinfo {author} {\bibfnamefont
  {H.-K.}\ \bibnamefont {Lo}}, \ and\ \bibinfo {author} {\bibfnamefont
  {I.}~\bibnamefont {Tzitrin}},\ }\bibfield  {title} {\enquote {\bibinfo
  {title} {{Quantum repeaters: From quantum networks to the quantum
  internet}},}\ }\href {\doibase 10.1103/RevModPhys.95.045006} {\bibfield
  {journal} {\bibinfo  {journal} {Reviews of Modern Physics}\ }\textbf
  {\bibinfo {volume} {95}},\ \bibinfo {pages} {045006} (\bibinfo {year}
  {2023})}\BibitemShut {NoStop}%
\bibitem [{\citenamefont {Langenfeld}\ \emph {et~al.}(2021)\citenamefont
  {Langenfeld}, \citenamefont {Thomas}, \citenamefont {Morin},\ and\
  \citenamefont {Rempe}}]{Langenfeld2021}%
  \BibitemOpen
  \bibfield  {author} {\bibinfo {author} {\bibfnamefont {S.}~\bibnamefont
  {Langenfeld}}, \bibinfo {author} {\bibfnamefont {P.}~\bibnamefont {Thomas}},
  \bibinfo {author} {\bibfnamefont {O.}~\bibnamefont {Morin}}, \ and\ \bibinfo
  {author} {\bibfnamefont {G.}~\bibnamefont {Rempe}},\ }\bibfield  {title}
  {\enquote {\bibinfo {title} {{Quantum Repeater Node Demonstrating
  Unconditionally Secure Key Distribution}},}\ }\href {\doibase
  10.1103/PhysRevLett.126.230506} {\bibfield  {journal} {\bibinfo  {journal}
  {Physical Review Letters}\ }\textbf {\bibinfo {volume} {126}},\ \bibinfo
  {pages} {230506} (\bibinfo {year} {2021})}\BibitemShut {NoStop}%
\bibitem [{\citenamefont {Jiang}\ \emph {et~al.}(2007)\citenamefont {Jiang},
  \citenamefont {Taylor}, \citenamefont {S{\o}rensen},\ and\ \citenamefont
  {Lukin}}]{Jiang2007}%
  \BibitemOpen
  \bibfield  {author} {\bibinfo {author} {\bibfnamefont {L.}~\bibnamefont
  {Jiang}}, \bibinfo {author} {\bibfnamefont {J.~M.}\ \bibnamefont {Taylor}},
  \bibinfo {author} {\bibfnamefont {A.~S.}\ \bibnamefont {S{\o}rensen}}, \ and\
  \bibinfo {author} {\bibfnamefont {M.~D.}\ \bibnamefont {Lukin}},\ }\bibfield
  {title} {\enquote {\bibinfo {title} {{Distributed quantum computation based
  on small quantum registers}},}\ }\href {\doibase 10.1103/PhysRevA.76.062323}
  {\bibfield  {journal} {\bibinfo  {journal} {Physical Review A}\ }\textbf
  {\bibinfo {volume} {76}},\ \bibinfo {pages} {062323} (\bibinfo {year}
  {2007})}\BibitemShut {NoStop}%
\bibitem [{\citenamefont {Malia}\ \emph {et~al.}(2022)\citenamefont {Malia},
  \citenamefont {Wu}, \citenamefont {Mart{\'{i}}nez-Rinc{\'{o}}n},\ and\
  \citenamefont {Kasevich}}]{Malia2022}%
  \BibitemOpen
  \bibfield  {author} {\bibinfo {author} {\bibfnamefont {B.~K.}\ \bibnamefont
  {Malia}}, \bibinfo {author} {\bibfnamefont {Y.}~\bibnamefont {Wu}}, \bibinfo
  {author} {\bibfnamefont {J.}~\bibnamefont {Mart{\'{i}}nez-Rinc{\'{o}}n}}, \
  and\ \bibinfo {author} {\bibfnamefont {M.~A.}\ \bibnamefont {Kasevich}},\
  }\bibfield  {title} {\enquote {\bibinfo {title} {{Distributed quantum sensing
  with mode-entangled spin-squeezed atomic states}},}\ }\href {\doibase
  10.1038/s41586-022-05363-z} {\bibfield  {journal} {\bibinfo  {journal}
  {Nature}\ }\textbf {\bibinfo {volume} {612}},\ \bibinfo {pages} {661}
  (\bibinfo {year} {2022})}\BibitemShut {NoStop}%
\bibitem [{\citenamefont {Guo}\ \emph {et~al.}(2020)\citenamefont {Guo},
  \citenamefont {Breum}, \citenamefont {Borregaard}, \citenamefont {Izumi},
  \citenamefont {Larsen}, \citenamefont {Gehring}, \citenamefont {Christandl},
  \citenamefont {Neergaard-Nielsen},\ and\ \citenamefont {Andersen}}]{Guo2020}%
  \BibitemOpen
  \bibfield  {author} {\bibinfo {author} {\bibfnamefont {X.}~\bibnamefont
  {Guo}}, \bibinfo {author} {\bibfnamefont {C.~R.}\ \bibnamefont {Breum}},
  \bibinfo {author} {\bibfnamefont {J.}~\bibnamefont {Borregaard}}, \bibinfo
  {author} {\bibfnamefont {S.}~\bibnamefont {Izumi}}, \bibinfo {author}
  {\bibfnamefont {M.~V.}\ \bibnamefont {Larsen}}, \bibinfo {author}
  {\bibfnamefont {T.}~\bibnamefont {Gehring}}, \bibinfo {author} {\bibfnamefont
  {M.}~\bibnamefont {Christandl}}, \bibinfo {author} {\bibfnamefont {J.~S.}\
  \bibnamefont {Neergaard-Nielsen}}, \ and\ \bibinfo {author} {\bibfnamefont
  {U.~L.}\ \bibnamefont {Andersen}},\ }\bibfield  {title} {\enquote {\bibinfo
  {title} {{Distributed quantum sensing in a continuous-variable entangled
  network}},}\ }\href {\doibase 10.1038/s41567-019-0743-x} {\bibfield
  {journal} {\bibinfo  {journal} {Nature Physics}\ }\textbf {\bibinfo {volume}
  {16}},\ \bibinfo {pages} {281} (\bibinfo {year} {2020})}\BibitemShut
  {NoStop}%
\bibitem [{\citenamefont {Chou}\ \emph {et~al.}(2005)\citenamefont {Chou},
  \citenamefont {De~Riedmatten}, \citenamefont {Felinto}, \citenamefont
  {Polyakov}, \citenamefont {Van~Enk},\ and\ \citenamefont
  {Kimble}}]{chou2005measurement}%
  \BibitemOpen
  \bibfield  {author} {\bibinfo {author} {\bibfnamefont {C.-W.}\ \bibnamefont
  {Chou}}, \bibinfo {author} {\bibfnamefont {H.}~\bibnamefont {De~Riedmatten}},
  \bibinfo {author} {\bibfnamefont {D.}~\bibnamefont {Felinto}}, \bibinfo
  {author} {\bibfnamefont {S.~V.}\ \bibnamefont {Polyakov}}, \bibinfo {author}
  {\bibfnamefont {S.~J.}\ \bibnamefont {Van~Enk}}, \ and\ \bibinfo {author}
  {\bibfnamefont {H.~J.}\ \bibnamefont {Kimble}},\ }\bibfield  {title}
  {\enquote {\bibinfo {title} {Measurement-induced entanglement for excitation
  stored in remote atomic ensembles},}\ }\href {\doibase
  10.1109/cleo.2006.4628575} {\bibfield  {journal} {\bibinfo  {journal}
  {Nature}\ }\textbf {\bibinfo {volume} {438}},\ \bibinfo {pages} {828}
  (\bibinfo {year} {2005})}\BibitemShut {NoStop}%
\bibitem [{\citenamefont {Hofmann}\ \emph {et~al.}(2012)\citenamefont
  {Hofmann}, \citenamefont {Krug}, \citenamefont {Ortegel}, \citenamefont
  {G{\'e}rard}, \citenamefont {Weber}, \citenamefont {Rosenfeld},\ and\
  \citenamefont {Weinfurter}}]{hofmann2012heralded}%
  \BibitemOpen
  \bibfield  {author} {\bibinfo {author} {\bibfnamefont {J.}~\bibnamefont
  {Hofmann}}, \bibinfo {author} {\bibfnamefont {M.}~\bibnamefont {Krug}},
  \bibinfo {author} {\bibfnamefont {N.}~\bibnamefont {Ortegel}}, \bibinfo
  {author} {\bibfnamefont {L.}~\bibnamefont {G{\'e}rard}}, \bibinfo {author}
  {\bibfnamefont {M.}~\bibnamefont {Weber}}, \bibinfo {author} {\bibfnamefont
  {W.}~\bibnamefont {Rosenfeld}}, \ and\ \bibinfo {author} {\bibfnamefont
  {H.}~\bibnamefont {Weinfurter}},\ }\bibfield  {title} {\enquote {\bibinfo
  {title} {Heralded entanglement between widely separated atoms},}\ }\href
  {https://www.science.org/doi/full/10.1126/science.1221856} {\bibfield
  {journal} {\bibinfo  {journal} {Science}\ }\textbf {\bibinfo {volume}
  {337}},\ \bibinfo {pages} {72} (\bibinfo {year} {2012})}\BibitemShut
  {NoStop}%
\bibitem [{\citenamefont {Moehring}\ \emph {et~al.}(2007)\citenamefont
  {Moehring}, \citenamefont {Maunz}, \citenamefont {Olmschenk}, \citenamefont
  {Younge}, \citenamefont {Matsukevich}, \citenamefont {Duan},\ and\
  \citenamefont {Monroe}}]{moehring2007entanglement}%
  \BibitemOpen
  \bibfield  {author} {\bibinfo {author} {\bibfnamefont {D.~L.}\ \bibnamefont
  {Moehring}}, \bibinfo {author} {\bibfnamefont {P.}~\bibnamefont {Maunz}},
  \bibinfo {author} {\bibfnamefont {S.}~\bibnamefont {Olmschenk}}, \bibinfo
  {author} {\bibfnamefont {K.~C.}\ \bibnamefont {Younge}}, \bibinfo {author}
  {\bibfnamefont {D.~N.}\ \bibnamefont {Matsukevich}}, \bibinfo {author}
  {\bibfnamefont {L.-M.}\ \bibnamefont {Duan}}, \ and\ \bibinfo {author}
  {\bibfnamefont {C.}~\bibnamefont {Monroe}},\ }\bibfield  {title} {\enquote
  {\bibinfo {title} {Entanglement of single-atom quantum bits at a distance},}\
  }\href {\doibase 10.1038/nature06118} {\bibfield  {journal} {\bibinfo
  {journal} {Nature}\ }\textbf {\bibinfo {volume} {449}},\ \bibinfo {pages}
  {68} (\bibinfo {year} {2007})}\BibitemShut {NoStop}%
\bibitem [{\citenamefont {Kurpiers}\ \emph {et~al.}(2018)\citenamefont
  {Kurpiers}, \citenamefont {Magnard}, \citenamefont {Walter}, \citenamefont
  {Royer}, \citenamefont {Pechal}, \citenamefont {Heinsoo}, \citenamefont
  {Salath{\'e}}, \citenamefont {Akin}, \citenamefont {Storz}, \citenamefont
  {Besse} \emph {et~al.}}]{kurpiers2018deterministic}%
  \BibitemOpen
  \bibfield  {author} {\bibinfo {author} {\bibfnamefont {P.}~\bibnamefont
  {Kurpiers}}, \bibinfo {author} {\bibfnamefont {P.}~\bibnamefont {Magnard}},
  \bibinfo {author} {\bibfnamefont {T.}~\bibnamefont {Walter}}, \bibinfo
  {author} {\bibfnamefont {B.}~\bibnamefont {Royer}}, \bibinfo {author}
  {\bibfnamefont {M.}~\bibnamefont {Pechal}}, \bibinfo {author} {\bibfnamefont
  {J.}~\bibnamefont {Heinsoo}}, \bibinfo {author} {\bibfnamefont
  {Y.}~\bibnamefont {Salath{\'e}}}, \bibinfo {author} {\bibfnamefont
  {A.}~\bibnamefont {Akin}}, \bibinfo {author} {\bibfnamefont {S.}~\bibnamefont
  {Storz}}, \bibinfo {author} {\bibfnamefont {J.-C.}\ \bibnamefont {Besse}},
  \emph {et~al.},\ }\bibfield  {title} {\enquote {\bibinfo {title}
  {Deterministic quantum state transfer and remote entanglement using microwave
  photons},}\ }\href {https://www.nature.com/articles/s41586-018-0195-y}
  {\bibfield  {journal} {\bibinfo  {journal} {Nature}\ }\textbf {\bibinfo
  {volume} {558}},\ \bibinfo {pages} {264} (\bibinfo {year}
  {2018})}\BibitemShut {NoStop}%
\bibitem [{\citenamefont {Delteil}\ \emph {et~al.}(2016)\citenamefont
  {Delteil}, \citenamefont {Sun}, \citenamefont {Gao}, \citenamefont {Togan},
  \citenamefont {Faelt},\ and\ \citenamefont
  {Imamo{\u{g}}lu}}]{delteil2016generation}%
  \BibitemOpen
  \bibfield  {author} {\bibinfo {author} {\bibfnamefont {A.}~\bibnamefont
  {Delteil}}, \bibinfo {author} {\bibfnamefont {Z.}~\bibnamefont {Sun}},
  \bibinfo {author} {\bibfnamefont {W.-b.}\ \bibnamefont {Gao}}, \bibinfo
  {author} {\bibfnamefont {E.}~\bibnamefont {Togan}}, \bibinfo {author}
  {\bibfnamefont {S.}~\bibnamefont {Faelt}}, \ and\ \bibinfo {author}
  {\bibfnamefont {A.}~\bibnamefont {Imamo{\u{g}}lu}},\ }\bibfield  {title}
  {\enquote {\bibinfo {title} {Generation of heralded entanglement between
  distant hole spins},}\ }\href {https://www.nature.com/articles/nphys3605}
  {\bibfield  {journal} {\bibinfo  {journal} {Nature Physics}\ }\textbf
  {\bibinfo {volume} {12}},\ \bibinfo {pages} {218} (\bibinfo {year}
  {2016})}\BibitemShut {NoStop}%
\bibitem [{\citenamefont {Humphreys}\ \emph {et~al.}(2018)\citenamefont
  {Humphreys}, \citenamefont {Kalb}, \citenamefont {Morits}, \citenamefont
  {Schouten}, \citenamefont {Vermeulen}, \citenamefont {Twitchen},
  \citenamefont {Markham},\ and\ \citenamefont
  {Hanson}}]{humphreys2018deterministic}%
  \BibitemOpen
  \bibfield  {author} {\bibinfo {author} {\bibfnamefont {P.~C.}\ \bibnamefont
  {Humphreys}}, \bibinfo {author} {\bibfnamefont {N.}~\bibnamefont {Kalb}},
  \bibinfo {author} {\bibfnamefont {J.~P.}\ \bibnamefont {Morits}}, \bibinfo
  {author} {\bibfnamefont {R.~N.}\ \bibnamefont {Schouten}}, \bibinfo {author}
  {\bibfnamefont {R.~F.}\ \bibnamefont {Vermeulen}}, \bibinfo {author}
  {\bibfnamefont {D.~J.}\ \bibnamefont {Twitchen}}, \bibinfo {author}
  {\bibfnamefont {M.}~\bibnamefont {Markham}}, \ and\ \bibinfo {author}
  {\bibfnamefont {R.}~\bibnamefont {Hanson}},\ }\bibfield  {title} {\enquote
  {\bibinfo {title} {Deterministic delivery of remote entanglement on a quantum
  network},}\ }\href {https://www.nature.com/articles/s41586-018-0200-5}
  {\bibfield  {journal} {\bibinfo  {journal} {Nature}\ }\textbf {\bibinfo
  {volume} {558}},\ \bibinfo {pages} {268} (\bibinfo {year}
  {2018})}\BibitemShut {NoStop}%
\bibitem [{\citenamefont {Yuan}\ \emph {et~al.}(2010)\citenamefont {Yuan},
  \citenamefont {Bao}, \citenamefont {Lu}, \citenamefont {Zhang}, \citenamefont
  {Peng},\ and\ \citenamefont {Pan}}]{yuan2010entangled}%
  \BibitemOpen
  \bibfield  {author} {\bibinfo {author} {\bibfnamefont {Z.-S.}\ \bibnamefont
  {Yuan}}, \bibinfo {author} {\bibfnamefont {X.-H.}\ \bibnamefont {Bao}},
  \bibinfo {author} {\bibfnamefont {C.-Y.}\ \bibnamefont {Lu}}, \bibinfo
  {author} {\bibfnamefont {J.}~\bibnamefont {Zhang}}, \bibinfo {author}
  {\bibfnamefont {C.-Z.}\ \bibnamefont {Peng}}, \ and\ \bibinfo {author}
  {\bibfnamefont {J.-W.}\ \bibnamefont {Pan}},\ }\bibfield  {title} {\enquote
  {\bibinfo {title} {Entangled photons and quantum communication},}\ }\href
  {https://www.sciencedirect.com/science/article/pii/S0370157310001833}
  {\bibfield  {journal} {\bibinfo  {journal} {Physics Reports}\ }\textbf
  {\bibinfo {volume} {497}},\ \bibinfo {pages} {1} (\bibinfo {year}
  {2010})}\BibitemShut {NoStop}%
\bibitem [{\citenamefont {Hensen}\ \emph {et~al.}(2015)\citenamefont {Hensen},
  \citenamefont {Bernien}, \citenamefont {Dr{\'e}au}, \citenamefont {Reiserer},
  \citenamefont {Kalb}, \citenamefont {Blok}, \citenamefont {Ruitenberg},
  \citenamefont {Vermeulen}, \citenamefont {Schouten}, \citenamefont
  {Abell{\'a}n} \emph {et~al.}}]{hensen2015loophole}%
  \BibitemOpen
  \bibfield  {author} {\bibinfo {author} {\bibfnamefont {B.}~\bibnamefont
  {Hensen}}, \bibinfo {author} {\bibfnamefont {H.}~\bibnamefont {Bernien}},
  \bibinfo {author} {\bibfnamefont {A.~E.}\ \bibnamefont {Dr{\'e}au}}, \bibinfo
  {author} {\bibfnamefont {A.}~\bibnamefont {Reiserer}}, \bibinfo {author}
  {\bibfnamefont {N.}~\bibnamefont {Kalb}}, \bibinfo {author} {\bibfnamefont
  {M.~S.}\ \bibnamefont {Blok}}, \bibinfo {author} {\bibfnamefont
  {J.}~\bibnamefont {Ruitenberg}}, \bibinfo {author} {\bibfnamefont {R.~F.}\
  \bibnamefont {Vermeulen}}, \bibinfo {author} {\bibfnamefont {R.~N.}\
  \bibnamefont {Schouten}}, \bibinfo {author} {\bibfnamefont {C.}~\bibnamefont
  {Abell{\'a}n}},  \emph {et~al.},\ }\bibfield  {title} {\enquote {\bibinfo
  {title} {Loophole-free bell inequality violation using electron spins
  separated by 1.3 kilometres},}\ }\href
  {https://www.nature.com/articles/nature15759} {\bibfield  {journal} {\bibinfo
   {journal} {Nature}\ }\textbf {\bibinfo {volume} {526}},\ \bibinfo {pages}
  {682} (\bibinfo {year} {2015})}\BibitemShut {NoStop}%
\bibitem [{\citenamefont {Zhou}\ \emph
  {et~al.}(2024{\natexlab{a}})\citenamefont {Zhou}, \citenamefont {Malik},
  \citenamefont {Fertig}, \citenamefont {Bock}, \citenamefont {Bauer},
  \citenamefont {van Leent}, \citenamefont {Zhang}, \citenamefont {Becher},\
  and\ \citenamefont {Weinfurter}}]{Zhou2024long}%
  \BibitemOpen
  \bibfield  {author} {\bibinfo {author} {\bibfnamefont {Y.}~\bibnamefont
  {Zhou}}, \bibinfo {author} {\bibfnamefont {P.}~\bibnamefont {Malik}},
  \bibinfo {author} {\bibfnamefont {F.}~\bibnamefont {Fertig}}, \bibinfo
  {author} {\bibfnamefont {M.}~\bibnamefont {Bock}}, \bibinfo {author}
  {\bibfnamefont {T.}~\bibnamefont {Bauer}}, \bibinfo {author} {\bibfnamefont
  {T.}~\bibnamefont {van Leent}}, \bibinfo {author} {\bibfnamefont
  {W.}~\bibnamefont {Zhang}}, \bibinfo {author} {\bibfnamefont
  {C.}~\bibnamefont {Becher}}, \ and\ \bibinfo {author} {\bibfnamefont
  {H.}~\bibnamefont {Weinfurter}},\ }\bibfield  {title} {\enquote {\bibinfo
  {title} {{Long-Lived Quantum Memory Enabling Atom-Photon Entanglement over
  101 km of Telecom Fiber}},}\ }\href {\doibase 10.1103/PRXQuantum.5.020307}
  {\bibfield  {journal} {\bibinfo  {journal} {PRX Quantum}\ }\textbf {\bibinfo
  {volume} {5}},\ \bibinfo {pages} {020307} (\bibinfo {year}
  {2024}{\natexlab{a}})}\BibitemShut {NoStop}%
\bibitem [{\citenamefont {Gonzalez-Tudela}\ \emph {et~al.}(2011)\citenamefont
  {Gonzalez-Tudela}, \citenamefont {Martin-Cano}, \citenamefont {Moreno},
  \citenamefont {Martin-Moreno}, \citenamefont {Tejedor},\ and\ \citenamefont
  {Garcia-Vidal}}]{gonzalez2011entanglement}%
  \BibitemOpen
  \bibfield  {author} {\bibinfo {author} {\bibfnamefont {A.}~\bibnamefont
  {Gonzalez-Tudela}}, \bibinfo {author} {\bibfnamefont {D.}~\bibnamefont
  {Martin-Cano}}, \bibinfo {author} {\bibfnamefont {E.}~\bibnamefont {Moreno}},
  \bibinfo {author} {\bibfnamefont {L.}~\bibnamefont {Martin-Moreno}}, \bibinfo
  {author} {\bibfnamefont {C.}~\bibnamefont {Tejedor}}, \ and\ \bibinfo
  {author} {\bibfnamefont {F.~J.}\ \bibnamefont {Garcia-Vidal}},\ }\bibfield
  {title} {\enquote {\bibinfo {title} {Entanglement of two qubits mediated by
  one-dimensional plasmonic waveguides},}\ }\href
  {https://journals.aps.org/prl/pdf/10.1103/PhysRevLett.106.020501} {\bibfield
  {journal} {\bibinfo  {journal} {Physical Review letters}\ }\textbf {\bibinfo
  {volume} {106}},\ \bibinfo {pages} {020501} (\bibinfo {year}
  {2011})}\BibitemShut {NoStop}%
\bibitem [{\citenamefont {Inagaki}\ \emph {et~al.}(2013)\citenamefont
  {Inagaki}, \citenamefont {Matsuda}, \citenamefont {Tadanaga}, \citenamefont
  {Asobe},\ and\ \citenamefont {Takesue}}]{inagaki2013entanglement}%
  \BibitemOpen
  \bibfield  {author} {\bibinfo {author} {\bibfnamefont {T.}~\bibnamefont
  {Inagaki}}, \bibinfo {author} {\bibfnamefont {N.}~\bibnamefont {Matsuda}},
  \bibinfo {author} {\bibfnamefont {O.}~\bibnamefont {Tadanaga}}, \bibinfo
  {author} {\bibfnamefont {M.}~\bibnamefont {Asobe}}, \ and\ \bibinfo {author}
  {\bibfnamefont {H.}~\bibnamefont {Takesue}},\ }\bibfield  {title} {\enquote
  {\bibinfo {title} {Entanglement distribution over 300 km of fiber},}\ }\href
  {https://opg.optica.org/oe/fulltext.cfm?uri=oe-21-20-23241&id=267617}
  {\bibfield  {journal} {\bibinfo  {journal} {Optics express}\ }\textbf
  {\bibinfo {volume} {21}},\ \bibinfo {pages} {23241} (\bibinfo {year}
  {2013})}\BibitemShut {NoStop}%
\bibitem [{\citenamefont {Yu}\ \emph {et~al.}(2020)\citenamefont {Yu},
  \citenamefont {Ma}, \citenamefont {Luo}, \citenamefont {Jing}, \citenamefont
  {Sun}, \citenamefont {Fang}, \citenamefont {Yang}, \citenamefont {Liu},
  \citenamefont {Zheng}, \citenamefont {Xie} \emph
  {et~al.}}]{yu2020entanglement}%
  \BibitemOpen
  \bibfield  {author} {\bibinfo {author} {\bibfnamefont {Y.}~\bibnamefont
  {Yu}}, \bibinfo {author} {\bibfnamefont {F.}~\bibnamefont {Ma}}, \bibinfo
  {author} {\bibfnamefont {X.-Y.}\ \bibnamefont {Luo}}, \bibinfo {author}
  {\bibfnamefont {B.}~\bibnamefont {Jing}}, \bibinfo {author} {\bibfnamefont
  {P.-F.}\ \bibnamefont {Sun}}, \bibinfo {author} {\bibfnamefont {R.-Z.}\
  \bibnamefont {Fang}}, \bibinfo {author} {\bibfnamefont {C.-W.}\ \bibnamefont
  {Yang}}, \bibinfo {author} {\bibfnamefont {H.}~\bibnamefont {Liu}}, \bibinfo
  {author} {\bibfnamefont {M.-Y.}\ \bibnamefont {Zheng}}, \bibinfo {author}
  {\bibfnamefont {X.-P.}\ \bibnamefont {Xie}},  \emph {et~al.},\ }\bibfield
  {title} {\enquote {\bibinfo {title} {Entanglement of two quantum memories via
  fibres over dozens of kilometres},}\ }\href
  {https://www.nature.com/articles/s41586-020-1976-7} {\bibfield  {journal}
  {\bibinfo  {journal} {Nature}\ }\textbf {\bibinfo {volume} {578}},\ \bibinfo
  {pages} {240} (\bibinfo {year} {2020})}\BibitemShut {NoStop}%
\bibitem [{\citenamefont {Yin}\ \emph {et~al.}(2017)\citenamefont {Yin},
  \citenamefont {Cao}, \citenamefont {Li}, \citenamefont {Liao}, \citenamefont
  {Zhang}, \citenamefont {Ren}, \citenamefont {Cai}, \citenamefont {Liu},
  \citenamefont {Li}, \citenamefont {Dai} \emph {et~al.}}]{yin2017satellite}%
  \BibitemOpen
  \bibfield  {author} {\bibinfo {author} {\bibfnamefont {J.}~\bibnamefont
  {Yin}}, \bibinfo {author} {\bibfnamefont {Y.}~\bibnamefont {Cao}}, \bibinfo
  {author} {\bibfnamefont {Y.-H.}\ \bibnamefont {Li}}, \bibinfo {author}
  {\bibfnamefont {S.-K.}\ \bibnamefont {Liao}}, \bibinfo {author}
  {\bibfnamefont {L.}~\bibnamefont {Zhang}}, \bibinfo {author} {\bibfnamefont
  {J.-G.}\ \bibnamefont {Ren}}, \bibinfo {author} {\bibfnamefont {W.-Q.}\
  \bibnamefont {Cai}}, \bibinfo {author} {\bibfnamefont {W.-Y.}\ \bibnamefont
  {Liu}}, \bibinfo {author} {\bibfnamefont {B.}~\bibnamefont {Li}}, \bibinfo
  {author} {\bibfnamefont {H.}~\bibnamefont {Dai}},  \emph {et~al.},\
  }\bibfield  {title} {\enquote {\bibinfo {title} {Satellite-based entanglement
  distribution over 1200 kilometers},}\ }\href
  {https://www.science.org/doi/full/10.1126/science.aan3211} {\bibfield
  {journal} {\bibinfo  {journal} {Science}\ }\textbf {\bibinfo {volume}
  {356}},\ \bibinfo {pages} {1140} (\bibinfo {year} {2017})}\BibitemShut
  {NoStop}%
\bibitem [{\citenamefont {Huang}\ \emph {et~al.}(2024)\citenamefont {Huang},
  \citenamefont {Salces-Carcoba}, \citenamefont {Adhikari}, \citenamefont
  {Safavi-Naeini},\ and\ \citenamefont {Jiang}}]{huang2024}%
  \BibitemOpen
  \bibfield  {author} {\bibinfo {author} {\bibfnamefont {Y.}~\bibnamefont
  {Huang}}, \bibinfo {author} {\bibfnamefont {F.}~\bibnamefont
  {Salces-Carcoba}}, \bibinfo {author} {\bibfnamefont {R.~X.}\ \bibnamefont
  {Adhikari}}, \bibinfo {author} {\bibfnamefont {A.~H.}\ \bibnamefont
  {Safavi-Naeini}}, \ and\ \bibinfo {author} {\bibfnamefont {L.}~\bibnamefont
  {Jiang}},\ }\bibfield  {title} {\enquote {\bibinfo {title} {Vacuum beam guide
  for large scale quantum networks},}\ }\href {\doibase
  10.1103/PhysRevLett.133.020801} {\bibfield  {journal} {\bibinfo  {journal}
  {Phys. Rev. Lett.}\ }\textbf {\bibinfo {volume} {133}},\ \bibinfo {pages}
  {020801} (\bibinfo {year} {2024})}\BibitemShut {NoStop}%
\bibitem [{\citenamefont {Kumar}(1990)}]{kumar1990quantum}%
  \BibitemOpen
  \bibfield  {author} {\bibinfo {author} {\bibfnamefont {P.}~\bibnamefont
  {Kumar}},\ }\bibfield  {title} {\enquote {\bibinfo {title} {Quantum frequency
  conversion},}\ }\href {\doibase 10.1364/oam.1990.fbb5} {\bibfield  {journal}
  {\bibinfo  {journal} {Optics letters}\ }\textbf {\bibinfo {volume} {15}},\
  \bibinfo {pages} {1476} (\bibinfo {year} {1990})}\BibitemShut {NoStop}%
\bibitem [{\citenamefont {Ikuta}\ \emph {et~al.}(2011)\citenamefont {Ikuta},
  \citenamefont {Kusaka}, \citenamefont {Kitano}, \citenamefont {Kato},
  \citenamefont {Yamamoto}, \citenamefont {Koashi},\ and\ \citenamefont
  {Imoto}}]{ikuta2011wide}%
  \BibitemOpen
  \bibfield  {author} {\bibinfo {author} {\bibfnamefont {R.}~\bibnamefont
  {Ikuta}}, \bibinfo {author} {\bibfnamefont {Y.}~\bibnamefont {Kusaka}},
  \bibinfo {author} {\bibfnamefont {T.}~\bibnamefont {Kitano}}, \bibinfo
  {author} {\bibfnamefont {H.}~\bibnamefont {Kato}}, \bibinfo {author}
  {\bibfnamefont {T.}~\bibnamefont {Yamamoto}}, \bibinfo {author}
  {\bibfnamefont {M.}~\bibnamefont {Koashi}}, \ and\ \bibinfo {author}
  {\bibfnamefont {N.}~\bibnamefont {Imoto}},\ }\bibfield  {title} {\enquote
  {\bibinfo {title} {Wide-band quantum interface for
  visible-to-telecommunication wavelength conversion},}\ }\href
  {https://www.nature.com/articles/ncomms1544} {\bibfield  {journal} {\bibinfo
  {journal} {Nature communications}\ }\textbf {\bibinfo {volume} {2}},\
  \bibinfo {pages} {537} (\bibinfo {year} {2011})}\BibitemShut {NoStop}%
\bibitem [{\citenamefont {Krutyanskiy}\ \emph {et~al.}(2024)\citenamefont
  {Krutyanskiy}, \citenamefont {Canteri}, \citenamefont {Meraner},
  \citenamefont {Krcmarsky},\ and\ \citenamefont {Lanyon}}]{Krutyanskiy2024}%
  \BibitemOpen
  \bibfield  {author} {\bibinfo {author} {\bibfnamefont {V.}~\bibnamefont
  {Krutyanskiy}}, \bibinfo {author} {\bibfnamefont {M.}~\bibnamefont
  {Canteri}}, \bibinfo {author} {\bibfnamefont {M.}~\bibnamefont {Meraner}},
  \bibinfo {author} {\bibfnamefont {V.}~\bibnamefont {Krcmarsky}}, \ and\
  \bibinfo {author} {\bibfnamefont {B.}~\bibnamefont {Lanyon}},\ }\bibfield
  {title} {\enquote {\bibinfo {title} {{Multimode Ion-Photon Entanglement over
  101 Kilometers}},}\ }\href {\doibase 10.1103/PRXQuantum.5.020308} {\bibfield
  {journal} {\bibinfo  {journal} {PRX Quantum}\ }\textbf {\bibinfo {volume}
  {5}},\ \bibinfo {pages} {020308} (\bibinfo {year} {2024})}\BibitemShut
  {NoStop}%
\bibitem [{\citenamefont {Zhou}\ \emph
  {et~al.}(2024{\natexlab{b}})\citenamefont {Zhou}, \citenamefont {Tamura},
  \citenamefont {Chang},\ and\ \citenamefont {Hung}}]{Zhou2024}%
  \BibitemOpen
  \bibfield  {author} {\bibinfo {author} {\bibfnamefont {X.}~\bibnamefont
  {Zhou}}, \bibinfo {author} {\bibfnamefont {H.}~\bibnamefont {Tamura}},
  \bibinfo {author} {\bibfnamefont {T.-H.}\ \bibnamefont {Chang}}, \ and\
  \bibinfo {author} {\bibfnamefont {C.-L.}\ \bibnamefont {Hung}},\ }\bibfield
  {title} {\enquote {\bibinfo {title} {{Trapped Atoms and Superradiance on an
  Integrated Nanophotonic Microring Circuit}},}\ }\href {\doibase
  10.1103/PhysRevX.14.031004} {\bibfield  {journal} {\bibinfo  {journal}
  {Physical Review X}\ }\textbf {\bibinfo {volume} {14}},\ \bibinfo {pages}
  {031004} (\bibinfo {year} {2024}{\natexlab{b}})}\BibitemShut {NoStop}%
\bibitem [{\citenamefont {Rabl}\ \emph {et~al.}(2010)\citenamefont {Rabl},
  \citenamefont {Kolkowitz}, \citenamefont {Koppens}, \citenamefont {Harris},
  \citenamefont {Zoller},\ and\ \citenamefont {Lukin}}]{rabl2010quantum}%
  \BibitemOpen
  \bibfield  {author} {\bibinfo {author} {\bibfnamefont {P.}~\bibnamefont
  {Rabl}}, \bibinfo {author} {\bibfnamefont {S.~J.}\ \bibnamefont {Kolkowitz}},
  \bibinfo {author} {\bibfnamefont {F.}~\bibnamefont {Koppens}}, \bibinfo
  {author} {\bibfnamefont {J.}~\bibnamefont {Harris}}, \bibinfo {author}
  {\bibfnamefont {P.}~\bibnamefont {Zoller}}, \ and\ \bibinfo {author}
  {\bibfnamefont {M.~D.}\ \bibnamefont {Lukin}},\ }\bibfield  {title} {\enquote
  {\bibinfo {title} {A quantum spin transducer based on nanoelectromechanical
  resonator arrays},}\ }\href {https://www.nature.com/articles/nphys1679}
  {\bibfield  {journal} {\bibinfo  {journal} {Nature Physics}\ }\textbf
  {\bibinfo {volume} {6}},\ \bibinfo {pages} {602} (\bibinfo {year}
  {2010})}\BibitemShut {NoStop}%
\bibitem [{\citenamefont {Benito}\ and\ \citenamefont
  {Burkard}(2020)}]{benito2020hybrid}%
  \BibitemOpen
  \bibfield  {author} {\bibinfo {author} {\bibfnamefont {M.}~\bibnamefont
  {Benito}}\ and\ \bibinfo {author} {\bibfnamefont {G.}~\bibnamefont
  {Burkard}},\ }\bibfield  {title} {\enquote {\bibinfo {title} {Hybrid
  superconductor-semiconductor systems for quantum technology},}\ }\href
  {https://pubs.aip.org/aip/apl/article/116/19/190502/1061270} {\bibfield
  {journal} {\bibinfo  {journal} {Applied Physics Letters}\ }\textbf {\bibinfo
  {volume} {116}},\ \bibinfo {pages} {190502} (\bibinfo {year}
  {2020})}\BibitemShut {NoStop}%
\bibitem [{\citenamefont {Mirhosseini}\ \emph {et~al.}(2020)\citenamefont
  {Mirhosseini}, \citenamefont {Sipahigil}, \citenamefont {Kalaee},\ and\
  \citenamefont {Painter}}]{mirhosseini2020superconducting}%
  \BibitemOpen
  \bibfield  {author} {\bibinfo {author} {\bibfnamefont {M.}~\bibnamefont
  {Mirhosseini}}, \bibinfo {author} {\bibfnamefont {A.}~\bibnamefont
  {Sipahigil}}, \bibinfo {author} {\bibfnamefont {M.}~\bibnamefont {Kalaee}}, \
  and\ \bibinfo {author} {\bibfnamefont {O.}~\bibnamefont {Painter}},\
  }\bibfield  {title} {\enquote {\bibinfo {title} {Superconducting qubit to
  optical photon transduction},}\ }\href
  {https://www.nature.com/articles/s41586-020-3038-6} {\bibfield  {journal}
  {\bibinfo  {journal} {Nature}\ }\textbf {\bibinfo {volume} {588}},\ \bibinfo
  {pages} {599} (\bibinfo {year} {2020})}\BibitemShut {NoStop}%
\bibitem [{\citenamefont {Wallquist}\ \emph {et~al.}(2009)\citenamefont
  {Wallquist}, \citenamefont {Hammerer}, \citenamefont {Rabl}, \citenamefont
  {Lukin},\ and\ \citenamefont {Zoller}}]{wallquist2009hybrid}%
  \BibitemOpen
  \bibfield  {author} {\bibinfo {author} {\bibfnamefont {M.}~\bibnamefont
  {Wallquist}}, \bibinfo {author} {\bibfnamefont {K.}~\bibnamefont {Hammerer}},
  \bibinfo {author} {\bibfnamefont {P.}~\bibnamefont {Rabl}}, \bibinfo {author}
  {\bibfnamefont {M.}~\bibnamefont {Lukin}}, \ and\ \bibinfo {author}
  {\bibfnamefont {P.}~\bibnamefont {Zoller}},\ }\bibfield  {title} {\enquote
  {\bibinfo {title} {Hybrid quantum devices and quantum engineering},}\ }\href
  {\doibase 10.1088/0031-8949/2009/t137/014001} {\bibfield  {journal} {\bibinfo
   {journal} {Physica Scripta}\ }\textbf {\bibinfo {volume} {2009}},\ \bibinfo
  {pages} {014001} (\bibinfo {year} {2009})}\BibitemShut {NoStop}%
\bibitem [{\citenamefont {Kurizki}\ \emph {et~al.}(2015)\citenamefont
  {Kurizki}, \citenamefont {Bertet}, \citenamefont {Kubo}, \citenamefont
  {M{\o}lmer}, \citenamefont {Petrosyan}, \citenamefont {Rabl},\ and\
  \citenamefont {Schmiedmayer}}]{kurizki2015quantum}%
  \BibitemOpen
  \bibfield  {author} {\bibinfo {author} {\bibfnamefont {G.}~\bibnamefont
  {Kurizki}}, \bibinfo {author} {\bibfnamefont {P.}~\bibnamefont {Bertet}},
  \bibinfo {author} {\bibfnamefont {Y.}~\bibnamefont {Kubo}}, \bibinfo {author}
  {\bibfnamefont {K.}~\bibnamefont {M{\o}lmer}}, \bibinfo {author}
  {\bibfnamefont {D.}~\bibnamefont {Petrosyan}}, \bibinfo {author}
  {\bibfnamefont {P.}~\bibnamefont {Rabl}}, \ and\ \bibinfo {author}
  {\bibfnamefont {J.}~\bibnamefont {Schmiedmayer}},\ }\bibfield  {title}
  {\enquote {\bibinfo {title} {Quantum technologies with hybrid systems},}\
  }\href {https://www.pnas.org/doi/abs/10.1073/pnas.1419326112} {\bibfield
  {journal} {\bibinfo  {journal} {Proceedings of the National Academy of
  Sciences}\ }\textbf {\bibinfo {volume} {112}},\ \bibinfo {pages} {3866}
  (\bibinfo {year} {2015})}\BibitemShut {NoStop}%
\bibitem [{\citenamefont {Han}\ \emph {et~al.}(2021)\citenamefont {Han},
  \citenamefont {Fu}, \citenamefont {Zou}, \citenamefont {Jiang},\ and\
  \citenamefont {Tang}}]{Han2021}%
  \BibitemOpen
  \bibfield  {author} {\bibinfo {author} {\bibfnamefont {X.}~\bibnamefont
  {Han}}, \bibinfo {author} {\bibfnamefont {W.}~\bibnamefont {Fu}}, \bibinfo
  {author} {\bibfnamefont {C.-L.}\ \bibnamefont {Zou}}, \bibinfo {author}
  {\bibfnamefont {L.}~\bibnamefont {Jiang}}, \ and\ \bibinfo {author}
  {\bibfnamefont {H.~X.}\ \bibnamefont {Tang}},\ }\bibfield  {title} {\enquote
  {\bibinfo {title} {{Microwave-optical quantum frequency conversion}},}\
  }\href {\doibase 10.1364/OPTICA.425414} {\bibfield  {journal} {\bibinfo
  {journal} {Optica}\ }\textbf {\bibinfo {volume} {8}},\ \bibinfo {pages}
  {1050} (\bibinfo {year} {2021})}\BibitemShut {NoStop}%
\bibitem [{\citenamefont {Balram}\ and\ \citenamefont
  {Srinivasan}(2022)}]{Balram2022}%
  \BibitemOpen
  \bibfield  {author} {\bibinfo {author} {\bibfnamefont {K.~C.}\ \bibnamefont
  {Balram}}\ and\ \bibinfo {author} {\bibfnamefont {K.}~\bibnamefont
  {Srinivasan}},\ }\bibfield  {title} {\enquote {\bibinfo {title}
  {Piezoelectric optomechanical approaches for efficient quantum
  microwave-to-optical signal transduction: the need for co-design},}\ }\href
  {\doibase 10.1002/qute.202100095} {\bibfield  {journal} {\bibinfo  {journal}
  {Advanced Quantum Technologies}\ }\textbf {\bibinfo {volume} {5}},\ \bibinfo
  {pages} {2100095} (\bibinfo {year} {2022})}\BibitemShut {NoStop}%
\bibitem [{\citenamefont {Liu}\ \emph {et~al.}(2024)\citenamefont {Liu},
  \citenamefont {Luo}, \citenamefont {Yu}, \citenamefont {Wang}, \citenamefont
  {Wang}, \citenamefont {Hu}, \citenamefont {Li}, \citenamefont {Zheng},
  \citenamefont {Yao}, \citenamefont {Yan} \emph {et~al.}}]{liu2024creation}%
  \BibitemOpen
  \bibfield  {author} {\bibinfo {author} {\bibfnamefont {J.-L.}\ \bibnamefont
  {Liu}}, \bibinfo {author} {\bibfnamefont {X.-Y.}\ \bibnamefont {Luo}},
  \bibinfo {author} {\bibfnamefont {Y.}~\bibnamefont {Yu}}, \bibinfo {author}
  {\bibfnamefont {C.-Y.}\ \bibnamefont {Wang}}, \bibinfo {author}
  {\bibfnamefont {B.}~\bibnamefont {Wang}}, \bibinfo {author} {\bibfnamefont
  {Y.}~\bibnamefont {Hu}}, \bibinfo {author} {\bibfnamefont {J.}~\bibnamefont
  {Li}}, \bibinfo {author} {\bibfnamefont {M.-Y.}\ \bibnamefont {Zheng}},
  \bibinfo {author} {\bibfnamefont {B.}~\bibnamefont {Yao}}, \bibinfo {author}
  {\bibfnamefont {Z.}~\bibnamefont {Yan}},  \emph {et~al.},\ }\bibfield
  {title} {\enquote {\bibinfo {title} {Creation of memory--memory entanglement
  in a metropolitan quantum network},}\ }\href
  {https://www.nature.com/articles/s41586-024-07308-0} {\bibfield  {journal}
  {\bibinfo  {journal} {Nature}\ }\textbf {\bibinfo {volume} {629}},\ \bibinfo
  {pages} {579} (\bibinfo {year} {2024})}\BibitemShut {NoStop}%
\bibitem [{\citenamefont {Knaut}\ \emph
  {et~al.}(2024{\natexlab{a}})\citenamefont {Knaut}, \citenamefont
  {Suleymanzade}, \citenamefont {Wei}, \citenamefont {Assumpcao}, \citenamefont
  {Stas}, \citenamefont {Huan}, \citenamefont {Machielse}, \citenamefont
  {Knall}, \citenamefont {Sutula}, \citenamefont {Baranes}, \citenamefont
  {Sinclair}, \citenamefont {De-Eknamkul}, \citenamefont {Levonian},
  \citenamefont {Bhaskar}, \citenamefont {Park}, \citenamefont {Lon{\v{c}}ar},\
  and\ \citenamefont {Lukin}}]{Knaut2024}%
  \BibitemOpen
  \bibfield  {author} {\bibinfo {author} {\bibfnamefont {C.~M.}\ \bibnamefont
  {Knaut}}, \bibinfo {author} {\bibfnamefont {A.}~\bibnamefont {Suleymanzade}},
  \bibinfo {author} {\bibfnamefont {Y.-C.}\ \bibnamefont {Wei}}, \bibinfo
  {author} {\bibfnamefont {D.~R.}\ \bibnamefont {Assumpcao}}, \bibinfo {author}
  {\bibfnamefont {P.-J.}\ \bibnamefont {Stas}}, \bibinfo {author}
  {\bibfnamefont {Y.~Q.}\ \bibnamefont {Huan}}, \bibinfo {author}
  {\bibfnamefont {B.}~\bibnamefont {Machielse}}, \bibinfo {author}
  {\bibfnamefont {E.~N.}\ \bibnamefont {Knall}}, \bibinfo {author}
  {\bibfnamefont {M.}~\bibnamefont {Sutula}}, \bibinfo {author} {\bibfnamefont
  {G.}~\bibnamefont {Baranes}}, \bibinfo {author} {\bibfnamefont
  {N.}~\bibnamefont {Sinclair}}, \bibinfo {author} {\bibfnamefont
  {C.}~\bibnamefont {De-Eknamkul}}, \bibinfo {author} {\bibfnamefont {D.~S.}\
  \bibnamefont {Levonian}}, \bibinfo {author} {\bibfnamefont {M.~K.}\
  \bibnamefont {Bhaskar}}, \bibinfo {author} {\bibfnamefont {H.}~\bibnamefont
  {Park}}, \bibinfo {author} {\bibfnamefont {M.}~\bibnamefont {Lon{\v{c}}ar}},
  \ and\ \bibinfo {author} {\bibfnamefont {M.~D.}\ \bibnamefont {Lukin}},\
  }\bibfield  {title} {\enquote {\bibinfo {title} {{Entanglement of
  nanophotonic quantum memory nodes in a telecom network}},}\ }\href {\doibase
  10.1038/s41586-024-07252-z} {\bibfield  {journal} {\bibinfo  {journal}
  {Nature}\ }\textbf {\bibinfo {volume} {629}},\ \bibinfo {pages} {573}
  (\bibinfo {year} {2024}{\natexlab{a}})}\BibitemShut {NoStop}%
\bibitem [{\citenamefont {Stolk}\ \emph {et~al.}(2024)\citenamefont {Stolk},
  \citenamefont {van~der Enden}, \citenamefont {Slater}, \citenamefont
  {te~Raa-Derckx}, \citenamefont {Botma}, \citenamefont {van Rantwijk},
  \citenamefont {Biemond}, \citenamefont {Hagen}, \citenamefont {Herfst},
  \citenamefont {Koek}, \citenamefont {Meskers}, \citenamefont {Vollmer},
  \citenamefont {van Zwet}, \citenamefont {Markham}, \citenamefont {Edmonds},
  \citenamefont {Geus}, \citenamefont {Elsen}, \citenamefont {Jungbluth},
  \citenamefont {Haefner}, \citenamefont {Tresp}, \citenamefont {Stuhler},
  \citenamefont {Ritter},\ and\ \citenamefont {Hanson}}]{Stolk2024}%
  \BibitemOpen
  \bibfield  {author} {\bibinfo {author} {\bibfnamefont {A.~J.}\ \bibnamefont
  {Stolk}}, \bibinfo {author} {\bibfnamefont {K.~L.}\ \bibnamefont {van~der
  Enden}}, \bibinfo {author} {\bibfnamefont {M.-C.}\ \bibnamefont {Slater}},
  \bibinfo {author} {\bibfnamefont {I.}~\bibnamefont {te~Raa-Derckx}}, \bibinfo
  {author} {\bibfnamefont {P.}~\bibnamefont {Botma}}, \bibinfo {author}
  {\bibfnamefont {J.}~\bibnamefont {van Rantwijk}}, \bibinfo {author}
  {\bibfnamefont {B.}~\bibnamefont {Biemond}}, \bibinfo {author} {\bibfnamefont
  {R.~A.~J.}\ \bibnamefont {Hagen}}, \bibinfo {author} {\bibfnamefont {R.~W.}\
  \bibnamefont {Herfst}}, \bibinfo {author} {\bibfnamefont {W.~D.}\
  \bibnamefont {Koek}}, \bibinfo {author} {\bibfnamefont {A.~J.~H.}\
  \bibnamefont {Meskers}}, \bibinfo {author} {\bibfnamefont {R.}~\bibnamefont
  {Vollmer}}, \bibinfo {author} {\bibfnamefont {E.~J.}\ \bibnamefont {van
  Zwet}}, \bibinfo {author} {\bibfnamefont {M.}~\bibnamefont {Markham}},
  \bibinfo {author} {\bibfnamefont {A.~M.}\ \bibnamefont {Edmonds}}, \bibinfo
  {author} {\bibfnamefont {J.~F.}\ \bibnamefont {Geus}}, \bibinfo {author}
  {\bibfnamefont {F.}~\bibnamefont {Elsen}}, \bibinfo {author} {\bibfnamefont
  {B.}~\bibnamefont {Jungbluth}}, \bibinfo {author} {\bibfnamefont
  {C.}~\bibnamefont {Haefner}}, \bibinfo {author} {\bibfnamefont
  {C.}~\bibnamefont {Tresp}}, \bibinfo {author} {\bibfnamefont
  {J.}~\bibnamefont {Stuhler}}, \bibinfo {author} {\bibfnamefont
  {S.}~\bibnamefont {Ritter}}, \ and\ \bibinfo {author} {\bibfnamefont
  {R.}~\bibnamefont {Hanson}},\ }\bibfield  {title} {\enquote {\bibinfo {title}
  {{Metropolitan-scale heralded entanglement of solid-state qubits}},}\ }\href
  {http://arxiv.org/abs/2404.03723} {\bibfield  {journal} {\bibinfo  {journal}
  {arXiv}\ ,\ \bibinfo {pages} {2404.03723}} (\bibinfo {year}
  {2024})}\BibitemShut {NoStop}%
\bibitem [{\citenamefont {van Leent}\ \emph {et~al.}(2022)\citenamefont {van
  Leent}, \citenamefont {Bock}, \citenamefont {Fertig}, \citenamefont
  {Garthoff}, \citenamefont {Eppelt}, \citenamefont {Zhou}, \citenamefont
  {Malik}, \citenamefont {Seubert}, \citenamefont {Bauer}, \citenamefont
  {Rosenfeld} \emph {et~al.}}]{van2022entangling}%
  \BibitemOpen
  \bibfield  {author} {\bibinfo {author} {\bibfnamefont {T.}~\bibnamefont {van
  Leent}}, \bibinfo {author} {\bibfnamefont {M.}~\bibnamefont {Bock}}, \bibinfo
  {author} {\bibfnamefont {F.}~\bibnamefont {Fertig}}, \bibinfo {author}
  {\bibfnamefont {R.}~\bibnamefont {Garthoff}}, \bibinfo {author}
  {\bibfnamefont {S.}~\bibnamefont {Eppelt}}, \bibinfo {author} {\bibfnamefont
  {Y.}~\bibnamefont {Zhou}}, \bibinfo {author} {\bibfnamefont {P.}~\bibnamefont
  {Malik}}, \bibinfo {author} {\bibfnamefont {M.}~\bibnamefont {Seubert}},
  \bibinfo {author} {\bibfnamefont {T.}~\bibnamefont {Bauer}}, \bibinfo
  {author} {\bibfnamefont {W.}~\bibnamefont {Rosenfeld}},  \emph {et~al.},\
  }\bibfield  {title} {\enquote {\bibinfo {title} {Entangling single atoms over
  33 km telecom fibre},}\ }\href
  {https://www.nature.com/articles/s41586-022-04764-4} {\bibfield  {journal}
  {\bibinfo  {journal} {Nature}\ }\textbf {\bibinfo {volume} {607}},\ \bibinfo
  {pages} {69} (\bibinfo {year} {2022})}\BibitemShut {NoStop}%
\bibitem [{See Supplemental Material for details on theoretical
  derivations()}]{SM}%
  \BibitemOpen
  See Supplemental Material for details on theoretical derivations,\ \href@noop
  {} {}\BibitemShut {NoStop}%
\bibitem [{\citenamefont {Wang}\ \emph {et~al.}(2021)\citenamefont {Wang},
  \citenamefont {Yang}, \citenamefont {Li}, \citenamefont {Hu}, \citenamefont
  {Surya}, \citenamefont {Xu}, \citenamefont {Dong}, \citenamefont {Guo},
  \citenamefont {Tang},\ and\ \citenamefont {Zou}}]{Wang2021}%
  \BibitemOpen
  \bibfield  {author} {\bibinfo {author} {\bibfnamefont {J.-Q.}\ \bibnamefont
  {Wang}}, \bibinfo {author} {\bibfnamefont {Y.-H.}\ \bibnamefont {Yang}},
  \bibinfo {author} {\bibfnamefont {M.}~\bibnamefont {Li}}, \bibinfo {author}
  {\bibfnamefont {X.-X.}\ \bibnamefont {Hu}}, \bibinfo {author} {\bibfnamefont
  {J.~B.}\ \bibnamefont {Surya}}, \bibinfo {author} {\bibfnamefont {X.-B.}\
  \bibnamefont {Xu}}, \bibinfo {author} {\bibfnamefont {C.-H.}\ \bibnamefont
  {Dong}}, \bibinfo {author} {\bibfnamefont {G.-C.}\ \bibnamefont {Guo}},
  \bibinfo {author} {\bibfnamefont {H.~X.}\ \bibnamefont {Tang}}, \ and\
  \bibinfo {author} {\bibfnamefont {C.-L.}\ \bibnamefont {Zou}},\ }\bibfield
  {title} {\enquote {\bibinfo {title} {{Efficient Frequency Conversion in a
  Degenerate chi(2) Microresonator}},}\ }\href {\doibase
  10.1103/PhysRevLett.126.133601} {\bibfield  {journal} {\bibinfo  {journal}
  {Physical Review Letters}\ }\textbf {\bibinfo {volume} {126}},\ \bibinfo
  {pages} {133601} (\bibinfo {year} {2021})}\BibitemShut {NoStop}%
\bibitem [{\citenamefont {Ilchenko}\ \emph {et~al.}(2004)\citenamefont
  {Ilchenko}, \citenamefont {Savchenkov}, \citenamefont {Matsko},\ and\
  \citenamefont {Maleki}}]{ilchenko2004nonlinear}%
  \BibitemOpen
  \bibfield  {author} {\bibinfo {author} {\bibfnamefont {V.~S.}\ \bibnamefont
  {Ilchenko}}, \bibinfo {author} {\bibfnamefont {A.~A.}\ \bibnamefont
  {Savchenkov}}, \bibinfo {author} {\bibfnamefont {A.~B.}\ \bibnamefont
  {Matsko}}, \ and\ \bibinfo {author} {\bibfnamefont {L.}~\bibnamefont
  {Maleki}},\ }\bibfield  {title} {\enquote {\bibinfo {title} {Nonlinear optics
  and crystalline whispering gallery mode cavities},}\ }\href {\doibase
  10.1103/physrevlett.92.043903} {\bibfield  {journal} {\bibinfo  {journal}
  {Physical Review letters}\ }\textbf {\bibinfo {volume} {92}},\ \bibinfo
  {pages} {043903} (\bibinfo {year} {2004})}\BibitemShut {NoStop}%
\bibitem [{\citenamefont {Javerzac-Galy}\ \emph {et~al.}(2016)\citenamefont
  {Javerzac-Galy}, \citenamefont {Plekhanov}, \citenamefont {Bernier},
  \citenamefont {Toth}, \citenamefont {Feofanov},\ and\ \citenamefont
  {Kippenberg}}]{Javerzac-Galy2016}%
  \BibitemOpen
  \bibfield  {author} {\bibinfo {author} {\bibfnamefont {C.}~\bibnamefont
  {Javerzac-Galy}}, \bibinfo {author} {\bibfnamefont {K.}~\bibnamefont
  {Plekhanov}}, \bibinfo {author} {\bibfnamefont {N.~R.}\ \bibnamefont
  {Bernier}}, \bibinfo {author} {\bibfnamefont {L.~D.}\ \bibnamefont {Toth}},
  \bibinfo {author} {\bibfnamefont {A.~K.}\ \bibnamefont {Feofanov}}, \ and\
  \bibinfo {author} {\bibfnamefont {T.~J.}\ \bibnamefont {Kippenberg}},\
  }\bibfield  {title} {\enquote {\bibinfo {title} {{On-chip
  microwave-to-optical quantum coherent converter based on a superconducting
  resonator coupled to an electro-optic microresonator}},}\ }\href {\doibase
  10.1103/PhysRevA.94.053815} {\bibfield  {journal} {\bibinfo  {journal}
  {Physical Review A}\ }\textbf {\bibinfo {volume} {94}},\ \bibinfo {pages}
  {053815} (\bibinfo {year} {2016})}\BibitemShut {NoStop}%
\bibitem [{\citenamefont {Fan}\ \emph {et~al.}(2018)\citenamefont {Fan},
  \citenamefont {Zou}, \citenamefont {Cheng}, \citenamefont {Guo},
  \citenamefont {Han}, \citenamefont {Gong}, \citenamefont {Wang},\ and\
  \citenamefont {Tang}}]{fan2018superconducting}%
  \BibitemOpen
  \bibfield  {author} {\bibinfo {author} {\bibfnamefont {L.}~\bibnamefont
  {Fan}}, \bibinfo {author} {\bibfnamefont {C.-L.}\ \bibnamefont {Zou}},
  \bibinfo {author} {\bibfnamefont {R.}~\bibnamefont {Cheng}}, \bibinfo
  {author} {\bibfnamefont {X.}~\bibnamefont {Guo}}, \bibinfo {author}
  {\bibfnamefont {X.}~\bibnamefont {Han}}, \bibinfo {author} {\bibfnamefont
  {Z.}~\bibnamefont {Gong}}, \bibinfo {author} {\bibfnamefont {S.}~\bibnamefont
  {Wang}}, \ and\ \bibinfo {author} {\bibfnamefont {H.~X.}\ \bibnamefont
  {Tang}},\ }\bibfield  {title} {\enquote {\bibinfo {title} {Superconducting
  cavity electro-optics: a platform for coherent photon conversion between
  superconducting and photonic circuits},}\ }\href
  {https://www.science.org/doi/full/10.1126/sciadv.aar4994} {\bibfield
  {journal} {\bibinfo  {journal} {Science advances}\ }\textbf {\bibinfo
  {volume} {4}},\ \bibinfo {pages} {eaar4994} (\bibinfo {year}
  {2018})}\BibitemShut {NoStop}%
\bibitem [{\citenamefont {Shen}\ and\ \citenamefont
  {Fan}(2005)}]{shen2005coherent}%
  \BibitemOpen
  \bibfield  {author} {\bibinfo {author} {\bibfnamefont {J.-T.}\ \bibnamefont
  {Shen}}\ and\ \bibinfo {author} {\bibfnamefont {S.}~\bibnamefont {Fan}},\
  }\bibfield  {title} {\enquote {\bibinfo {title} {Coherent single photon
  transport in a one-dimensional waveguide coupled with superconducting quantum
  bits},}\ }\href {\doibase 10.1103/physrevlett.95.213001} {\bibfield
  {journal} {\bibinfo  {journal} {Physical Review letters}\ }\textbf {\bibinfo
  {volume} {95}},\ \bibinfo {pages} {213001} (\bibinfo {year}
  {2005})}\BibitemShut {NoStop}%
\bibitem [{\citenamefont {Rosenblum}\ \emph {et~al.}(2017)\citenamefont
  {Rosenblum}, \citenamefont {Borne},\ and\ \citenamefont
  {Dayan}}]{rosenblum2017analysis}%
  \BibitemOpen
  \bibfield  {author} {\bibinfo {author} {\bibfnamefont {S.}~\bibnamefont
  {Rosenblum}}, \bibinfo {author} {\bibfnamefont {A.}~\bibnamefont {Borne}}, \
  and\ \bibinfo {author} {\bibfnamefont {B.}~\bibnamefont {Dayan}},\ }\bibfield
   {title} {\enquote {\bibinfo {title} {Analysis of deterministic swapping of
  photonic and atomic states through single-photon raman interaction},}\ }\href
  {https://journals.aps.org/pra/abstract/10.1103/PhysRevA.95.033814} {\bibfield
   {journal} {\bibinfo  {journal} {Physical Review A}\ }\textbf {\bibinfo
  {volume} {95}},\ \bibinfo {pages} {033814} (\bibinfo {year}
  {2017})}\BibitemShut {NoStop}%
\bibitem [{\citenamefont {Yan}\ \emph {et~al.}(2023)\citenamefont {Yan},
  \citenamefont {Li}, \citenamefont {Xu}, \citenamefont {Zhang}, \citenamefont
  {Ma},\ and\ \citenamefont {Zou}}]{yan2023unidirectional}%
  \BibitemOpen
  \bibfield  {author} {\bibinfo {author} {\bibfnamefont {C.-H.}\ \bibnamefont
  {Yan}}, \bibinfo {author} {\bibfnamefont {M.}~\bibnamefont {Li}}, \bibinfo
  {author} {\bibfnamefont {X.-B.}\ \bibnamefont {Xu}}, \bibinfo {author}
  {\bibfnamefont {Y.-L.}\ \bibnamefont {Zhang}}, \bibinfo {author}
  {\bibfnamefont {X.-Y.}\ \bibnamefont {Ma}}, \ and\ \bibinfo {author}
  {\bibfnamefont {C.-L.}\ \bibnamefont {Zou}},\ }\bibfield  {title} {\enquote
  {\bibinfo {title} {Unidirectional propagation of single photons realized by a
  scatterer coupled to whispering-gallery-mode microresonators},}\ }\href
  {https://journals.aps.org/pra/abstract/10.1103/PhysRevA.107.033713}
  {\bibfield  {journal} {\bibinfo  {journal} {Physical Review A}\ }\textbf
  {\bibinfo {volume} {107}},\ \bibinfo {pages} {033713} (\bibinfo {year}
  {2023})}\BibitemShut {NoStop}%
\bibitem [{\citenamefont {Pan}\ \emph {et~al.}(1998)\citenamefont {Pan},
  \citenamefont {Bouwmeester}, \citenamefont {Weinfurter},\ and\ \citenamefont
  {Zeilinger}}]{pan1998experimental}%
  \BibitemOpen
  \bibfield  {author} {\bibinfo {author} {\bibfnamefont {J.-W.}\ \bibnamefont
  {Pan}}, \bibinfo {author} {\bibfnamefont {D.}~\bibnamefont {Bouwmeester}},
  \bibinfo {author} {\bibfnamefont {H.}~\bibnamefont {Weinfurter}}, \ and\
  \bibinfo {author} {\bibfnamefont {A.}~\bibnamefont {Zeilinger}},\ }\bibfield
  {title} {\enquote {\bibinfo {title} {Experimental entanglement swapping:
  entangling photons that never interacted},}\ }\href {\doibase
  10.1103/physrevlett.80.3891} {\bibfield  {journal} {\bibinfo  {journal}
  {Physical Review letters}\ }\textbf {\bibinfo {volume} {80}},\ \bibinfo
  {pages} {3891} (\bibinfo {year} {1998})}\BibitemShut {NoStop}%
\bibitem [{\citenamefont {Duan}\ and\ \citenamefont {Kimble}(2004)}]{Duan2004}%
  \BibitemOpen
  \bibfield  {author} {\bibinfo {author} {\bibfnamefont {L.-M.}\ \bibnamefont
  {Duan}}\ and\ \bibinfo {author} {\bibfnamefont {H.}~\bibnamefont {Kimble}},\
  }\bibfield  {title} {\enquote {\bibinfo {title} {{Scalable Photonic Quantum
  Computation through Cavity-Assisted Interactions}},}\ }\href {\doibase
  10.1103/physrevlett.92.127902} {\bibfield  {journal} {\bibinfo  {journal}
  {Physical Review Letters}\ }\textbf {\bibinfo {volume} {92}},\ \bibinfo
  {pages} {127902} (\bibinfo {year} {2004})}\BibitemShut {NoStop}%
\bibitem [{\citenamefont {Hofmann}\ and\ \citenamefont
  {Takeuchi}(2002)}]{hofmann2002quantum}%
  \BibitemOpen
  \bibfield  {author} {\bibinfo {author} {\bibfnamefont {H.~F.}\ \bibnamefont
  {Hofmann}}\ and\ \bibinfo {author} {\bibfnamefont {S.}~\bibnamefont
  {Takeuchi}},\ }\bibfield  {title} {\enquote {\bibinfo {title} {Quantum phase
  gate for photonic qubits using only beam splitters and postselection},}\
  }\href {\doibase 10.1103/physreva.66.024308} {\bibfield  {journal} {\bibinfo
  {journal} {Physical Review A}\ }\textbf {\bibinfo {volume} {66}},\ \bibinfo
  {pages} {024308} (\bibinfo {year} {2002})}\BibitemShut {NoStop}%
\bibitem [{\citenamefont {Christandl}\ \emph {et~al.}(2009)\citenamefont
  {Christandl}, \citenamefont {K{\"o}nig},\ and\ \citenamefont
  {Renner}}]{christandl2009postselection}%
  \BibitemOpen
  \bibfield  {author} {\bibinfo {author} {\bibfnamefont {M.}~\bibnamefont
  {Christandl}}, \bibinfo {author} {\bibfnamefont {R.}~\bibnamefont
  {K{\"o}nig}}, \ and\ \bibinfo {author} {\bibfnamefont {R.}~\bibnamefont
  {Renner}},\ }\bibfield  {title} {\enquote {\bibinfo {title} {Postselection
  technique for quantum channels with applications to quantum cryptography},}\
  }\href {\doibase 10.1103/physrevlett.102.020504} {\bibfield  {journal}
  {\bibinfo  {journal} {Physical Review letters}\ }\textbf {\bibinfo {volume}
  {102}},\ \bibinfo {pages} {020504} (\bibinfo {year} {2009})}\BibitemShut
  {NoStop}%
\bibitem [{\citenamefont {Arvidsson-Shukur}\ \emph {et~al.}(2020)\citenamefont
  {Arvidsson-Shukur}, \citenamefont {Yunger~Halpern}, \citenamefont {Lepage},
  \citenamefont {Lasek}, \citenamefont {Barnes},\ and\ \citenamefont
  {Lloyd}}]{arvidsson2020quantum}%
  \BibitemOpen
  \bibfield  {author} {\bibinfo {author} {\bibfnamefont {D.~R.}\ \bibnamefont
  {Arvidsson-Shukur}}, \bibinfo {author} {\bibfnamefont {N.}~\bibnamefont
  {Yunger~Halpern}}, \bibinfo {author} {\bibfnamefont {H.~V.}\ \bibnamefont
  {Lepage}}, \bibinfo {author} {\bibfnamefont {A.~A.}\ \bibnamefont {Lasek}},
  \bibinfo {author} {\bibfnamefont {C.~H.}\ \bibnamefont {Barnes}}, \ and\
  \bibinfo {author} {\bibfnamefont {S.}~\bibnamefont {Lloyd}},\ }\bibfield
  {title} {\enquote {\bibinfo {title} {Quantum advantage in postselected
  metrology},}\ }\href {https://www.nature.com/articles/s41467-020-17559-w}
  {\bibfield  {journal} {\bibinfo  {journal} {Nature communications}\ }\textbf
  {\bibinfo {volume} {11}},\ \bibinfo {pages} {3775} (\bibinfo {year}
  {2020})}\BibitemShut {NoStop}%
\bibitem [{\citenamefont {Xiang}\ \emph {et~al.}(2013)\citenamefont {Xiang},
  \citenamefont {Ashhab}, \citenamefont {You},\ and\ \citenamefont
  {Nori}}]{xiang2013hybrid}%
  \BibitemOpen
  \bibfield  {author} {\bibinfo {author} {\bibfnamefont {Z.-L.}\ \bibnamefont
  {Xiang}}, \bibinfo {author} {\bibfnamefont {S.}~\bibnamefont {Ashhab}},
  \bibinfo {author} {\bibfnamefont {J.}~\bibnamefont {You}}, \ and\ \bibinfo
  {author} {\bibfnamefont {F.}~\bibnamefont {Nori}},\ }\bibfield  {title}
  {\enquote {\bibinfo {title} {Hybrid quantum circuits: Superconducting
  circuits interacting with other quantum systems},}\ }\href
  {https://journals.aps.org/rmp/abstract/10.1103/RevModPhys.85.623} {\bibfield
  {journal} {\bibinfo  {journal} {Reviews of Modern Physics}\ }\textbf
  {\bibinfo {volume} {85}},\ \bibinfo {pages} {623} (\bibinfo {year}
  {2013})}\BibitemShut {NoStop}%
\bibitem [{\citenamefont {Clerk}\ \emph {et~al.}(2020)\citenamefont {Clerk},
  \citenamefont {Lehnert}, \citenamefont {Bertet}, \citenamefont {Petta},\ and\
  \citenamefont {Nakamura}}]{clerk2020hybrid}%
  \BibitemOpen
  \bibfield  {author} {\bibinfo {author} {\bibfnamefont {A.}~\bibnamefont
  {Clerk}}, \bibinfo {author} {\bibfnamefont {K.}~\bibnamefont {Lehnert}},
  \bibinfo {author} {\bibfnamefont {P.}~\bibnamefont {Bertet}}, \bibinfo
  {author} {\bibfnamefont {J.}~\bibnamefont {Petta}}, \ and\ \bibinfo {author}
  {\bibfnamefont {Y.}~\bibnamefont {Nakamura}},\ }\bibfield  {title} {\enquote
  {\bibinfo {title} {Hybrid quantum systems with circuit quantum
  electrodynamics},}\ }\href
  {https://www.nature.com/articles/s41567-020-0797-9} {\bibfield  {journal}
  {\bibinfo  {journal} {Nature Physics}\ }\textbf {\bibinfo {volume} {16}},\
  \bibinfo {pages} {257} (\bibinfo {year} {2020})}\BibitemShut {NoStop}%
\bibitem [{\citenamefont {Wang}\ \emph {et~al.}(2022)\citenamefont {Wang},
  \citenamefont {Zhang},\ and\ \citenamefont {Jiang}}]{wang2022generalized}%
  \BibitemOpen
  \bibfield  {author} {\bibinfo {author} {\bibfnamefont {C.-H.}\ \bibnamefont
  {Wang}}, \bibinfo {author} {\bibfnamefont {M.}~\bibnamefont {Zhang}}, \ and\
  \bibinfo {author} {\bibfnamefont {L.}~\bibnamefont {Jiang}},\ }\bibfield
  {title} {\enquote {\bibinfo {title} {Generalized matching condition for unity
  efficiency quantum transduction},}\ }\href
  {https://journals.aps.org/prresearch/abstract/10.1103/PhysRevResearch.4.L042023}
  {\bibfield  {journal} {\bibinfo  {journal} {Physical Review Research}\
  }\textbf {\bibinfo {volume} {4}},\ \bibinfo {pages} {L042023} (\bibinfo
  {year} {2022})}\BibitemShut {NoStop}%
\bibitem [{\citenamefont {Wang}\ and\ \citenamefont
  {Clerk}(2012)}]{wang2012using}%
  \BibitemOpen
  \bibfield  {author} {\bibinfo {author} {\bibfnamefont {Y.-D.}\ \bibnamefont
  {Wang}}\ and\ \bibinfo {author} {\bibfnamefont {A.~A.}\ \bibnamefont
  {Clerk}},\ }\bibfield  {title} {\enquote {\bibinfo {title} {Using dark modes
  for high-fidelity optomechanical quantum state transfer},}\ }\href
  {https://iopscience.iop.org/article/10.1088/1367-2630/14/10/105010}
  {\bibfield  {journal} {\bibinfo  {journal} {New Journal of Physics}\ }\textbf
  {\bibinfo {volume} {14}},\ \bibinfo {pages} {105010} (\bibinfo {year}
  {2012})}\BibitemShut {NoStop}%
\bibitem [{\citenamefont {Gao}\ \emph {et~al.}(2021)\citenamefont {Gao},
  \citenamefont {Zhang}, \citenamefont {Bo}, \citenamefont {Fang},
  \citenamefont {Hao}, \citenamefont {Yao}, \citenamefont {Lin}, \citenamefont
  {Guan}, \citenamefont {Deng}, \citenamefont {Wang} \emph
  {et~al.}}]{gao2021broadband}%
  \BibitemOpen
  \bibfield  {author} {\bibinfo {author} {\bibfnamefont {R.}~\bibnamefont
  {Gao}}, \bibinfo {author} {\bibfnamefont {H.}~\bibnamefont {Zhang}}, \bibinfo
  {author} {\bibfnamefont {F.}~\bibnamefont {Bo}}, \bibinfo {author}
  {\bibfnamefont {W.}~\bibnamefont {Fang}}, \bibinfo {author} {\bibfnamefont
  {Z.}~\bibnamefont {Hao}}, \bibinfo {author} {\bibfnamefont {N.}~\bibnamefont
  {Yao}}, \bibinfo {author} {\bibfnamefont {J.}~\bibnamefont {Lin}}, \bibinfo
  {author} {\bibfnamefont {J.}~\bibnamefont {Guan}}, \bibinfo {author}
  {\bibfnamefont {L.}~\bibnamefont {Deng}}, \bibinfo {author} {\bibfnamefont
  {M.}~\bibnamefont {Wang}},  \emph {et~al.},\ }\bibfield  {title} {\enquote
  {\bibinfo {title} {Broadband highly efficient nonlinear optical processes in
  on-chip integrated lithium niobate microdisk resonators of q-factor above
  108},}\ }\href
  {https://iopscience.iop.org/article/10.1088/1367-2630/ac3d52/meta} {\bibfield
   {journal} {\bibinfo  {journal} {New Journal of Physics}\ }\textbf {\bibinfo
  {volume} {23}},\ \bibinfo {pages} {123027} (\bibinfo {year}
  {2021})}\BibitemShut {NoStop}%
\bibitem [{\citenamefont {Lu}\ \emph {et~al.}(2020)\citenamefont {Lu},
  \citenamefont {Li}, \citenamefont {Zou}, \citenamefont {Al~Sayem},\ and\
  \citenamefont {Tang}}]{lu2020toward}%
  \BibitemOpen
  \bibfield  {author} {\bibinfo {author} {\bibfnamefont {J.}~\bibnamefont
  {Lu}}, \bibinfo {author} {\bibfnamefont {M.}~\bibnamefont {Li}}, \bibinfo
  {author} {\bibfnamefont {C.-L.}\ \bibnamefont {Zou}}, \bibinfo {author}
  {\bibfnamefont {A.}~\bibnamefont {Al~Sayem}}, \ and\ \bibinfo {author}
  {\bibfnamefont {H.~X.}\ \bibnamefont {Tang}},\ }\bibfield  {title} {\enquote
  {\bibinfo {title} {Toward 1\% single-photon anharmonicity with periodically
  poled lithium niobate microring resonators},}\ }\href
  {https://opg.optica.org/optica/fulltext.cfm?uri=optica-7-12-1654&id=442855}
  {\bibfield  {journal} {\bibinfo  {journal} {Optica}\ }\textbf {\bibinfo
  {volume} {7}},\ \bibinfo {pages} {1654} (\bibinfo {year} {2020})}\BibitemShut
  {NoStop}%
\bibitem [{\citenamefont {Vernon}\ \emph {et~al.}(2016)\citenamefont {Vernon},
  \citenamefont {Liscidini},\ and\ \citenamefont {Sipe}}]{Vernon2016}%
  \BibitemOpen
  \bibfield  {author} {\bibinfo {author} {\bibfnamefont {Z.}~\bibnamefont
  {Vernon}}, \bibinfo {author} {\bibfnamefont {M.}~\bibnamefont {Liscidini}}, \
  and\ \bibinfo {author} {\bibfnamefont {J.~E.}\ \bibnamefont {Sipe}},\
  }\bibfield  {title} {\enquote {\bibinfo {title} {{Quantum frequency
  conversion and strong coupling of photonic modes using four-wave mixing in
  integrated microresonators}},}\ }\href {\doibase 10.1103/PhysRevA.94.023810}
  {\bibfield  {journal} {\bibinfo  {journal} {Physical Review A}\ }\textbf
  {\bibinfo {volume} {94}},\ \bibinfo {pages} {023810} (\bibinfo {year}
  {2016})}\BibitemShut {NoStop}%
\bibitem [{\citenamefont {Shen}\ \emph {et~al.}(2023)\citenamefont {Shen},
  \citenamefont {Zhang}, \citenamefont {Chen}, \citenamefont {Xiao},
  \citenamefont {Zou}, \citenamefont {Guo},\ and\ \citenamefont
  {Dong}}]{shen2023nonreciprocal}%
  \BibitemOpen
  \bibfield  {author} {\bibinfo {author} {\bibfnamefont {Z.}~\bibnamefont
  {Shen}}, \bibinfo {author} {\bibfnamefont {Y.-L.}\ \bibnamefont {Zhang}},
  \bibinfo {author} {\bibfnamefont {Y.}~\bibnamefont {Chen}}, \bibinfo {author}
  {\bibfnamefont {Y.-F.}\ \bibnamefont {Xiao}}, \bibinfo {author}
  {\bibfnamefont {C.-L.}\ \bibnamefont {Zou}}, \bibinfo {author} {\bibfnamefont
  {G.-C.}\ \bibnamefont {Guo}}, \ and\ \bibinfo {author} {\bibfnamefont
  {C.-H.}\ \bibnamefont {Dong}},\ }\bibfield  {title} {\enquote {\bibinfo
  {title} {Nonreciprocal frequency conversion and mode routing in a
  microresonator},}\ }\href
  {https://journals.aps.org/prl/pdf/10.1103/PhysRevLett.130.013601} {\bibfield
  {journal} {\bibinfo  {journal} {Physical Review Letters}\ }\textbf {\bibinfo
  {volume} {130}},\ \bibinfo {pages} {013601} (\bibinfo {year}
  {2023})}\BibitemShut {NoStop}%
\bibitem [{\citenamefont {Lu}\ \emph {et~al.}(2021)\citenamefont {Lu},
  \citenamefont {Al~Sayem}, \citenamefont {Gong}, \citenamefont {Surya},
  \citenamefont {Zou},\ and\ \citenamefont {Tang}}]{lu2021ultralow}%
  \BibitemOpen
  \bibfield  {author} {\bibinfo {author} {\bibfnamefont {J.}~\bibnamefont
  {Lu}}, \bibinfo {author} {\bibfnamefont {A.}~\bibnamefont {Al~Sayem}},
  \bibinfo {author} {\bibfnamefont {Z.}~\bibnamefont {Gong}}, \bibinfo {author}
  {\bibfnamefont {J.~B.}\ \bibnamefont {Surya}}, \bibinfo {author}
  {\bibfnamefont {C.-L.}\ \bibnamefont {Zou}}, \ and\ \bibinfo {author}
  {\bibfnamefont {H.~X.}\ \bibnamefont {Tang}},\ }\bibfield  {title} {\enquote
  {\bibinfo {title} {Ultralow-threshold thin-film lithium niobate optical
  parametric oscillator},}\ }\href
  {https://opg.optica.org/optica/fulltext.cfm?uri=optica-8-4-539&id=450023}
  {\bibfield  {journal} {\bibinfo  {journal} {Optica}\ }\textbf {\bibinfo
  {volume} {8}},\ \bibinfo {pages} {539} (\bibinfo {year} {2021})}\BibitemShut
  {NoStop}%
\bibitem [{\citenamefont {Marty}\ \emph {et~al.}(2021)\citenamefont {Marty},
  \citenamefont {Combri{\'e}}, \citenamefont {Raineri},\ and\ \citenamefont
  {De~Rossi}}]{marty2021photonic}%
  \BibitemOpen
  \bibfield  {author} {\bibinfo {author} {\bibfnamefont {G.}~\bibnamefont
  {Marty}}, \bibinfo {author} {\bibfnamefont {S.}~\bibnamefont {Combri{\'e}}},
  \bibinfo {author} {\bibfnamefont {F.}~\bibnamefont {Raineri}}, \ and\
  \bibinfo {author} {\bibfnamefont {A.}~\bibnamefont {De~Rossi}},\ }\bibfield
  {title} {\enquote {\bibinfo {title} {Photonic crystal optical parametric
  oscillator},}\ }\href {https://www.nature.com/articles/s41566-020-00737-z}
  {\bibfield  {journal} {\bibinfo  {journal} {Nature photonics}\ }\textbf
  {\bibinfo {volume} {15}},\ \bibinfo {pages} {53} (\bibinfo {year}
  {2021})}\BibitemShut {NoStop}%
\bibitem [{\citenamefont {Tiecke}\ \emph {et~al.}(2014)\citenamefont {Tiecke},
  \citenamefont {Thompson}, \citenamefont {de~Leon}, \citenamefont {Liu},
  \citenamefont {Vuleti{\'c}},\ and\ \citenamefont
  {Lukin}}]{tiecke2014nanophotonic}%
  \BibitemOpen
  \bibfield  {author} {\bibinfo {author} {\bibfnamefont {T.}~\bibnamefont
  {Tiecke}}, \bibinfo {author} {\bibfnamefont {J.~D.}\ \bibnamefont
  {Thompson}}, \bibinfo {author} {\bibfnamefont {N.~P.}\ \bibnamefont
  {de~Leon}}, \bibinfo {author} {\bibfnamefont {L.}~\bibnamefont {Liu}},
  \bibinfo {author} {\bibfnamefont {V.}~\bibnamefont {Vuleti{\'c}}}, \ and\
  \bibinfo {author} {\bibfnamefont {M.~D.}\ \bibnamefont {Lukin}},\ }\bibfield
  {title} {\enquote {\bibinfo {title} {Nanophotonic quantum phase switch with a
  single atom},}\ }\href {https://www.nature.com/articles/nature13188}
  {\bibfield  {journal} {\bibinfo  {journal} {Nature}\ }\textbf {\bibinfo
  {volume} {508}},\ \bibinfo {pages} {241} (\bibinfo {year}
  {2014})}\BibitemShut {NoStop}%
\bibitem [{\citenamefont {Aoki}\ \emph {et~al.}(2006)\citenamefont {Aoki},
  \citenamefont {Dayan}, \citenamefont {Wilcut}, \citenamefont {Bowen},
  \citenamefont {Parkins}, \citenamefont {Kippenberg}, \citenamefont {Vahala},\
  and\ \citenamefont {Kimble}}]{aoki2006observation}%
  \BibitemOpen
  \bibfield  {author} {\bibinfo {author} {\bibfnamefont {T.}~\bibnamefont
  {Aoki}}, \bibinfo {author} {\bibfnamefont {B.}~\bibnamefont {Dayan}},
  \bibinfo {author} {\bibfnamefont {E.}~\bibnamefont {Wilcut}}, \bibinfo
  {author} {\bibfnamefont {W.~P.}\ \bibnamefont {Bowen}}, \bibinfo {author}
  {\bibfnamefont {A.~S.}\ \bibnamefont {Parkins}}, \bibinfo {author}
  {\bibfnamefont {T.}~\bibnamefont {Kippenberg}}, \bibinfo {author}
  {\bibfnamefont {K.}~\bibnamefont {Vahala}}, \ and\ \bibinfo {author}
  {\bibfnamefont {H.}~\bibnamefont {Kimble}},\ }\bibfield  {title} {\enquote
  {\bibinfo {title} {Observation of strong coupling between one atom and a
  monolithic microresonator},}\ }\href {\doibase 10.1038/nature05147}
  {\bibfield  {journal} {\bibinfo  {journal} {Nature}\ }\textbf {\bibinfo
  {volume} {443}},\ \bibinfo {pages} {671} (\bibinfo {year}
  {2006})}\BibitemShut {NoStop}%
\bibitem [{\citenamefont {Shomroni}\ \emph {et~al.}(2014)\citenamefont
  {Shomroni}, \citenamefont {Rosenblum}, \citenamefont {Lovsky}, \citenamefont
  {Bechler}, \citenamefont {Guendelman},\ and\ \citenamefont
  {Dayan}}]{shomroni2014all}%
  \BibitemOpen
  \bibfield  {author} {\bibinfo {author} {\bibfnamefont {I.}~\bibnamefont
  {Shomroni}}, \bibinfo {author} {\bibfnamefont {S.}~\bibnamefont {Rosenblum}},
  \bibinfo {author} {\bibfnamefont {Y.}~\bibnamefont {Lovsky}}, \bibinfo
  {author} {\bibfnamefont {O.}~\bibnamefont {Bechler}}, \bibinfo {author}
  {\bibfnamefont {G.}~\bibnamefont {Guendelman}}, \ and\ \bibinfo {author}
  {\bibfnamefont {B.}~\bibnamefont {Dayan}},\ }\bibfield  {title} {\enquote
  {\bibinfo {title} {All-optical routing of single photons by a one-atom switch
  controlled by a single photon},}\ }\href
  {https://www.science.org/doi/full/10.1126/science.1254699} {\bibfield
  {journal} {\bibinfo  {journal} {Science}\ }\textbf {\bibinfo {volume}
  {345}},\ \bibinfo {pages} {903} (\bibinfo {year} {2014})}\BibitemShut
  {NoStop}%
\bibitem [{\citenamefont {Scheucher}\ \emph {et~al.}(2016)\citenamefont
  {Scheucher}, \citenamefont {Hilico}, \citenamefont {Will}, \citenamefont
  {Volz},\ and\ \citenamefont {Rauschenbeutel}}]{scheucher2016quantum}%
  \BibitemOpen
  \bibfield  {author} {\bibinfo {author} {\bibfnamefont {M.}~\bibnamefont
  {Scheucher}}, \bibinfo {author} {\bibfnamefont {A.}~\bibnamefont {Hilico}},
  \bibinfo {author} {\bibfnamefont {E.}~\bibnamefont {Will}}, \bibinfo {author}
  {\bibfnamefont {J.}~\bibnamefont {Volz}}, \ and\ \bibinfo {author}
  {\bibfnamefont {A.}~\bibnamefont {Rauschenbeutel}},\ }\bibfield  {title}
  {\enquote {\bibinfo {title} {Quantum optical circulator controlled by a
  single chirally coupled atom},}\ }\href
  {https://www.science.org/doi/full/10.1126/science.aaj2118} {\bibfield
  {journal} {\bibinfo  {journal} {Science}\ }\textbf {\bibinfo {volume}
  {354}},\ \bibinfo {pages} {1577} (\bibinfo {year} {2016})}\BibitemShut
  {NoStop}%
\bibitem [{\citenamefont {Panuski}\ \emph {et~al.}(2020)\citenamefont
  {Panuski}, \citenamefont {Englund},\ and\ \citenamefont
  {Hamerly}}]{panuski2020fundamental}%
  \BibitemOpen
  \bibfield  {author} {\bibinfo {author} {\bibfnamefont {C.}~\bibnamefont
  {Panuski}}, \bibinfo {author} {\bibfnamefont {D.}~\bibnamefont {Englund}}, \
  and\ \bibinfo {author} {\bibfnamefont {R.}~\bibnamefont {Hamerly}},\
  }\bibfield  {title} {\enquote {\bibinfo {title} {Fundamental thermal noise
  limits for optical microcavities},}\ }\href
  {http://link.aps.org/pdf/10.1103/PhysRevX.10.041046} {\bibfield  {journal}
  {\bibinfo  {journal} {Physical Review X}\ }\textbf {\bibinfo {volume} {10}},\
  \bibinfo {pages} {041046} (\bibinfo {year} {2020})}\BibitemShut {NoStop}%
\bibitem [{\citenamefont {Han}\ \emph {et~al.}(2020)\citenamefont {Han},
  \citenamefont {Fu}, \citenamefont {Zhong}, \citenamefont {Zou}, \citenamefont
  {Xu}, \citenamefont {Sayem}, \citenamefont {Xu}, \citenamefont {Wang},
  \citenamefont {Cheng}, \citenamefont {Jiang} \emph {et~al.}}]{han2020cavity}%
  \BibitemOpen
  \bibfield  {author} {\bibinfo {author} {\bibfnamefont {X.}~\bibnamefont
  {Han}}, \bibinfo {author} {\bibfnamefont {W.}~\bibnamefont {Fu}}, \bibinfo
  {author} {\bibfnamefont {C.}~\bibnamefont {Zhong}}, \bibinfo {author}
  {\bibfnamefont {C.-L.}\ \bibnamefont {Zou}}, \bibinfo {author} {\bibfnamefont
  {Y.}~\bibnamefont {Xu}}, \bibinfo {author} {\bibfnamefont {A.~A.}\
  \bibnamefont {Sayem}}, \bibinfo {author} {\bibfnamefont {M.}~\bibnamefont
  {Xu}}, \bibinfo {author} {\bibfnamefont {S.}~\bibnamefont {Wang}}, \bibinfo
  {author} {\bibfnamefont {R.}~\bibnamefont {Cheng}}, \bibinfo {author}
  {\bibfnamefont {L.}~\bibnamefont {Jiang}},  \emph {et~al.},\ }\bibfield
  {title} {\enquote {\bibinfo {title} {Cavity piezo-mechanics for
  superconducting-nanophotonic quantum interface},}\ }\href
  {https://www.nature.com/articles/s41467-020-17053-3} {\bibfield  {journal}
  {\bibinfo  {journal} {Nature communications}\ }\textbf {\bibinfo {volume}
  {11}},\ \bibinfo {pages} {3237} (\bibinfo {year} {2020})}\BibitemShut
  {NoStop}%
\bibitem [{\citenamefont {Li}\ \emph {et~al.}(2018)\citenamefont {Li},
  \citenamefont {Zhu},\ and\ \citenamefont {Agarwal}}]{li2018magnon}%
  \BibitemOpen
  \bibfield  {author} {\bibinfo {author} {\bibfnamefont {J.}~\bibnamefont
  {Li}}, \bibinfo {author} {\bibfnamefont {S.-Y.}\ \bibnamefont {Zhu}}, \ and\
  \bibinfo {author} {\bibfnamefont {G.}~\bibnamefont {Agarwal}},\ }\bibfield
  {title} {\enquote {\bibinfo {title} {Magnon-photon-phonon entanglement in
  cavity magnomechanics},}\ }\href
  {https://journals.aps.org/prl/pdf/10.1103/PhysRevLett.121.203601} {\bibfield
  {journal} {\bibinfo  {journal} {Physical Review Letters}\ }\textbf {\bibinfo
  {volume} {121}},\ \bibinfo {pages} {203601} (\bibinfo {year}
  {2018})}\BibitemShut {NoStop}%
\bibitem [{\citenamefont {Lachance-Quirion}\ \emph {et~al.}(2019)\citenamefont
  {Lachance-Quirion}, \citenamefont {Tabuchi}, \citenamefont {Gloppe},
  \citenamefont {Usami},\ and\ \citenamefont {Nakamura}}]{lachance2019hybrid}%
  \BibitemOpen
  \bibfield  {author} {\bibinfo {author} {\bibfnamefont {D.}~\bibnamefont
  {Lachance-Quirion}}, \bibinfo {author} {\bibfnamefont {Y.}~\bibnamefont
  {Tabuchi}}, \bibinfo {author} {\bibfnamefont {A.}~\bibnamefont {Gloppe}},
  \bibinfo {author} {\bibfnamefont {K.}~\bibnamefont {Usami}}, \ and\ \bibinfo
  {author} {\bibfnamefont {Y.}~\bibnamefont {Nakamura}},\ }\bibfield  {title}
  {\enquote {\bibinfo {title} {Hybrid quantum systems based on magnonics},}\
  }\href {https://iopscience.iop.org/article/10.7567/1882-0786/ab248d/meta}
  {\bibfield  {journal} {\bibinfo  {journal} {Applied Physics Express}\
  }\textbf {\bibinfo {volume} {12}},\ \bibinfo {pages} {070101} (\bibinfo
  {year} {2019})}\BibitemShut {NoStop}%
\bibitem [{\citenamefont {Yeo}\ \emph {et~al.}(2014)\citenamefont {Yeo},
  \citenamefont {de~Assis}, \citenamefont {Gloppe}, \citenamefont
  {Dupont-Ferrier}, \citenamefont {Verlot}, \citenamefont {Malik},
  \citenamefont {Dupuy}, \citenamefont {Claudon}, \citenamefont {G{\'e}rard},
  \citenamefont {Auff{\`e}ves} \emph {et~al.}}]{yeo2014strain}%
  \BibitemOpen
  \bibfield  {author} {\bibinfo {author} {\bibfnamefont {I.}~\bibnamefont
  {Yeo}}, \bibinfo {author} {\bibfnamefont {P.-L.}\ \bibnamefont {de~Assis}},
  \bibinfo {author} {\bibfnamefont {A.}~\bibnamefont {Gloppe}}, \bibinfo
  {author} {\bibfnamefont {E.}~\bibnamefont {Dupont-Ferrier}}, \bibinfo
  {author} {\bibfnamefont {P.}~\bibnamefont {Verlot}}, \bibinfo {author}
  {\bibfnamefont {N.~S.}\ \bibnamefont {Malik}}, \bibinfo {author}
  {\bibfnamefont {E.}~\bibnamefont {Dupuy}}, \bibinfo {author} {\bibfnamefont
  {J.}~\bibnamefont {Claudon}}, \bibinfo {author} {\bibfnamefont {J.-M.}\
  \bibnamefont {G{\'e}rard}}, \bibinfo {author} {\bibfnamefont
  {A.}~\bibnamefont {Auff{\`e}ves}},  \emph {et~al.},\ }\bibfield  {title}
  {\enquote {\bibinfo {title} {Strain-mediated coupling in a quantum
  dot--mechanical oscillator hybrid system},}\ }\href
  {https://www.nature.com/articles/nnano.2013.274} {\bibfield  {journal}
  {\bibinfo  {journal} {Nature nanotechnology}\ }\textbf {\bibinfo {volume}
  {9}},\ \bibinfo {pages} {106} (\bibinfo {year} {2014})}\BibitemShut {NoStop}%
\bibitem [{\citenamefont {Li}\ \emph {et~al.}(2020)\citenamefont {Li},
  \citenamefont {Zhou}, \citenamefont {Gao},\ and\ \citenamefont
  {Nori}}]{li2020enhancing}%
  \BibitemOpen
  \bibfield  {author} {\bibinfo {author} {\bibfnamefont {P.-B.}\ \bibnamefont
  {Li}}, \bibinfo {author} {\bibfnamefont {Y.}~\bibnamefont {Zhou}}, \bibinfo
  {author} {\bibfnamefont {W.-B.}\ \bibnamefont {Gao}}, \ and\ \bibinfo
  {author} {\bibfnamefont {F.}~\bibnamefont {Nori}},\ }\bibfield  {title}
  {\enquote {\bibinfo {title} {Enhancing spin-phonon and spin-spin interactions
  using linear resources in a hybrid quantum system},}\ }\href
  {https://journals.aps.org/prl/pdf/10.1103/PhysRevLett.125.153602} {\bibfield
  {journal} {\bibinfo  {journal} {Physical Review Letters}\ }\textbf {\bibinfo
  {volume} {125}},\ \bibinfo {pages} {153602} (\bibinfo {year}
  {2020})}\BibitemShut {NoStop}%
\bibitem [{\citenamefont {Yu}\ \emph {et~al.}(2023)\citenamefont {Yu},
  \citenamefont {Liu}, \citenamefont {Lee}, \citenamefont {Michler},
  \citenamefont {Reitzenstein}, \citenamefont {Srinivasan}, \citenamefont
  {Waks},\ and\ \citenamefont {Liu}}]{yu2023telecom}%
  \BibitemOpen
  \bibfield  {author} {\bibinfo {author} {\bibfnamefont {Y.}~\bibnamefont
  {Yu}}, \bibinfo {author} {\bibfnamefont {S.}~\bibnamefont {Liu}}, \bibinfo
  {author} {\bibfnamefont {C.-M.}\ \bibnamefont {Lee}}, \bibinfo {author}
  {\bibfnamefont {P.}~\bibnamefont {Michler}}, \bibinfo {author} {\bibfnamefont
  {S.}~\bibnamefont {Reitzenstein}}, \bibinfo {author} {\bibfnamefont
  {K.}~\bibnamefont {Srinivasan}}, \bibinfo {author} {\bibfnamefont
  {E.}~\bibnamefont {Waks}}, \ and\ \bibinfo {author} {\bibfnamefont
  {J.}~\bibnamefont {Liu}},\ }\bibfield  {title} {\enquote {\bibinfo {title}
  {Telecom-band quantum dot technologies for long-distance quantum networks},}\
  }\href {https://www.nature.com/articles/s41565-023-01528-7} {\bibfield
  {journal} {\bibinfo  {journal} {Nature Nanotechnology}\ }\textbf {\bibinfo
  {volume} {18}},\ \bibinfo {pages} {1389} (\bibinfo {year}
  {2023})}\BibitemShut {NoStop}%
\bibitem [{\citenamefont {Lai}\ \emph {et~al.}(2024)\citenamefont {Lai},
  \citenamefont {Wang}, \citenamefont {Shi}, \citenamefont {Cui}, \citenamefont
  {Wang}, \citenamefont {Zhang}, \citenamefont {Liu}, \citenamefont {Tian},
  \citenamefont {Sun}, \citenamefont {Chang} \emph
  {et~al.}}]{lai2024realization}%
  \BibitemOpen
  \bibfield  {author} {\bibinfo {author} {\bibfnamefont {P.-C.}\ \bibnamefont
  {Lai}}, \bibinfo {author} {\bibfnamefont {Y.}~\bibnamefont {Wang}}, \bibinfo
  {author} {\bibfnamefont {J.-X.}\ \bibnamefont {Shi}}, \bibinfo {author}
  {\bibfnamefont {Z.-B.}\ \bibnamefont {Cui}}, \bibinfo {author} {\bibfnamefont
  {Z.-Q.}\ \bibnamefont {Wang}}, \bibinfo {author} {\bibfnamefont
  {S.}~\bibnamefont {Zhang}}, \bibinfo {author} {\bibfnamefont {P.-Y.}\
  \bibnamefont {Liu}}, \bibinfo {author} {\bibfnamefont {Z.-C.}\ \bibnamefont
  {Tian}}, \bibinfo {author} {\bibfnamefont {Y.-D.}\ \bibnamefont {Sun}},
  \bibinfo {author} {\bibfnamefont {X.-Y.}\ \bibnamefont {Chang}},  \emph
  {et~al.},\ }\bibfield  {title} {\enquote {\bibinfo {title} {Realization of a
  crosstalk-free multi-ion node for long-distance quantum networking},}\ }\href
  {http://arxiv.org/pdf/2405.13369v1} {\bibfield  {journal} {\bibinfo
  {journal} {arXiv preprint arXiv:2405.13369}\ } (\bibinfo {year}
  {2024})}\BibitemShut {NoStop}%
\bibitem [{\citenamefont {Knaut}\ \emph
  {et~al.}(2024{\natexlab{b}})\citenamefont {Knaut}, \citenamefont
  {Suleymanzade}, \citenamefont {Wei}, \citenamefont {Assumpcao}, \citenamefont
  {Stas}, \citenamefont {Huan}, \citenamefont {Machielse}, \citenamefont
  {Knall}, \citenamefont {Sutula}, \citenamefont {Baranes} \emph
  {et~al.}}]{knaut2024entanglement}%
  \BibitemOpen
  \bibfield  {author} {\bibinfo {author} {\bibfnamefont {C.}~\bibnamefont
  {Knaut}}, \bibinfo {author} {\bibfnamefont {A.}~\bibnamefont {Suleymanzade}},
  \bibinfo {author} {\bibfnamefont {Y.-C.}\ \bibnamefont {Wei}}, \bibinfo
  {author} {\bibfnamefont {D.}~\bibnamefont {Assumpcao}}, \bibinfo {author}
  {\bibfnamefont {P.-J.}\ \bibnamefont {Stas}}, \bibinfo {author}
  {\bibfnamefont {Y.}~\bibnamefont {Huan}}, \bibinfo {author} {\bibfnamefont
  {B.}~\bibnamefont {Machielse}}, \bibinfo {author} {\bibfnamefont
  {E.}~\bibnamefont {Knall}}, \bibinfo {author} {\bibfnamefont
  {M.}~\bibnamefont {Sutula}}, \bibinfo {author} {\bibfnamefont
  {G.}~\bibnamefont {Baranes}},  \emph {et~al.},\ }\bibfield  {title} {\enquote
  {\bibinfo {title} {Entanglement of nanophotonic quantum memory nodes in a
  telecom network},}\ }\href
  {https://www.nature.com/articles/s41586-024-07252-z} {\bibfield  {journal}
  {\bibinfo  {journal} {Nature}\ }\textbf {\bibinfo {volume} {629}},\ \bibinfo
  {pages} {573} (\bibinfo {year} {2024}{\natexlab{b}})}\BibitemShut {NoStop}%
\bibitem [{\citenamefont {Chang}\ \emph {et~al.}(2020)\citenamefont {Chang},
  \citenamefont {Zhong}, \citenamefont {Bienfait}, \citenamefont {Chou},
  \citenamefont {Conner}, \citenamefont {Dumur}, \citenamefont {Grebel},
  \citenamefont {Peairs}, \citenamefont {Povey}, \citenamefont {Satzinger}
  \emph {et~al.}}]{chang2020remote}%
  \BibitemOpen
  \bibfield  {author} {\bibinfo {author} {\bibfnamefont {H.-S.}\ \bibnamefont
  {Chang}}, \bibinfo {author} {\bibfnamefont {Y.}~\bibnamefont {Zhong}},
  \bibinfo {author} {\bibfnamefont {A.}~\bibnamefont {Bienfait}}, \bibinfo
  {author} {\bibfnamefont {M.-H.}\ \bibnamefont {Chou}}, \bibinfo {author}
  {\bibfnamefont {C.~R.}\ \bibnamefont {Conner}}, \bibinfo {author}
  {\bibfnamefont {{\'E}.}~\bibnamefont {Dumur}}, \bibinfo {author}
  {\bibfnamefont {J.}~\bibnamefont {Grebel}}, \bibinfo {author} {\bibfnamefont
  {G.~A.}\ \bibnamefont {Peairs}}, \bibinfo {author} {\bibfnamefont {R.~G.}\
  \bibnamefont {Povey}}, \bibinfo {author} {\bibfnamefont {K.~J.}\ \bibnamefont
  {Satzinger}},  \emph {et~al.},\ }\bibfield  {title} {\enquote {\bibinfo
  {title} {Remote entanglement via adiabatic passage using a tunably
  dissipative quantum communication system},}\ }\href
  {https://journals.aps.org/prl/pdf/10.1103/PhysRevLett.124.240502} {\bibfield
  {journal} {\bibinfo  {journal} {Physical Review Letters}\ }\textbf {\bibinfo
  {volume} {124}},\ \bibinfo {pages} {240502} (\bibinfo {year}
  {2020})}\BibitemShut {NoStop}%
\bibitem [{\citenamefont {Bienfait}\ \emph {et~al.}(2019)\citenamefont
  {Bienfait}, \citenamefont {Satzinger}, \citenamefont {Zhong}, \citenamefont
  {Chang}, \citenamefont {Chou}, \citenamefont {Conner}, \citenamefont {Dumur},
  \citenamefont {Grebel}, \citenamefont {Peairs}, \citenamefont {Povey} \emph
  {et~al.}}]{bienfait2019phonon}%
  \BibitemOpen
  \bibfield  {author} {\bibinfo {author} {\bibfnamefont {A.}~\bibnamefont
  {Bienfait}}, \bibinfo {author} {\bibfnamefont {K.~J.}\ \bibnamefont
  {Satzinger}}, \bibinfo {author} {\bibfnamefont {Y.}~\bibnamefont {Zhong}},
  \bibinfo {author} {\bibfnamefont {H.-S.}\ \bibnamefont {Chang}}, \bibinfo
  {author} {\bibfnamefont {M.-H.}\ \bibnamefont {Chou}}, \bibinfo {author}
  {\bibfnamefont {C.~R.}\ \bibnamefont {Conner}}, \bibinfo {author}
  {\bibfnamefont {{\'E}.}~\bibnamefont {Dumur}}, \bibinfo {author}
  {\bibfnamefont {J.}~\bibnamefont {Grebel}}, \bibinfo {author} {\bibfnamefont
  {G.~A.}\ \bibnamefont {Peairs}}, \bibinfo {author} {\bibfnamefont {R.~G.}\
  \bibnamefont {Povey}},  \emph {et~al.},\ }\bibfield  {title} {\enquote
  {\bibinfo {title} {Phonon-mediated quantum state transfer and remote qubit
  entanglement},}\ }\href
  {https://www.science.org/doi/full/10.1126/science.aaw8415} {\bibfield
  {journal} {\bibinfo  {journal} {Science}\ }\textbf {\bibinfo {volume}
  {364}},\ \bibinfo {pages} {368} (\bibinfo {year} {2019})}\BibitemShut
  {NoStop}%
\bibitem [{\citenamefont {Yurke}\ \emph {et~al.}(1986)\citenamefont {Yurke},
  \citenamefont {McCall},\ and\ \citenamefont {Klauder}}]{yurke19862}%
  \BibitemOpen
  \bibfield  {author} {\bibinfo {author} {\bibfnamefont {B.}~\bibnamefont
  {Yurke}}, \bibinfo {author} {\bibfnamefont {S.~L.}\ \bibnamefont {McCall}}, \
  and\ \bibinfo {author} {\bibfnamefont {J.~R.}\ \bibnamefont {Klauder}},\
  }\bibfield  {title} {\enquote {\bibinfo {title} {Su (2) and su (1, 1)
  interferometers},}\ }\href {\doibase 10.1007/978-94-009-3051-3_43} {\bibfield
   {journal} {\bibinfo  {journal} {Physical Review A}\ }\textbf {\bibinfo
  {volume} {33}},\ \bibinfo {pages} {4033} (\bibinfo {year}
  {1986})}\BibitemShut {NoStop}%
\end{thebibliography}%
\end{document}